\pdfoutput=1
\documentclass[%
 reprint,
 amsmath,amssymb,longbibliography,
 aps,
 prb,
]{revtex4-2}
\usepackage{hyperref}
\usepackage{multirow}
\usepackage{booktabs}
\usepackage[dvipsnames]{xcolor}
\usepackage{dsfont}
\usepackage{nicefrac}
\usepackage{bm}
\usepackage{graphicx}
\usepackage{amsmath}

\begin{document}

\title{Inter-orbital nematicity and the origin of a single electron Fermi pocket in FeSe}

\author{Daniel Steffensen$^1$, Andreas Kreisel$^2$, P. J. Hirschfeld$^{1,3}$, Brian M. Andersen$^1$}
\affiliation{%
$^1$Niels Bohr Institute, University of Copenhagen, Jagtvej 128, DK-2200 Copenhagen, Denmark\\
$^2$Institut f\" ur Theoretische Physik, Universit\"at Leipzig, D-04103 Leipzig, Germany\\
$^3$Department of Physics, University of Florida, Gainesville, Florida 32611, USA}

\date{\today}% It is always \today, today,
             %  but any date may be 

\begin{abstract}
The electronic structure of the enigmatic iron-based superconductor FeSe has puzzled researchers since spectroscopic probes failed to observe the expected electron pocket at the $Y$ point in the 1-Fe Brillouin zone. It has been speculated that this pocket, essential for an understanding of the superconducting state, is either absent or incoherent. Here,
we perform a theoretical study of the preferred nematic order originating from nearest-neighbor Coulomb interactions in an electronic model relevant for FeSe. We find that at low temperatures the dominating nematic components are of inter-orbital $d_{xz}-d_{xy}$ and $d_{yz}-d_{xy}$ character, with spontaneously broken amplitudes for these two components. This inter-orbital nematic order naturally leads to distinct hybridization gaps at the $X$ and $Y$ points of the 1-Fe Brillouin zone, and may thereby produce highly anisotropic Fermi surfaces with only a single electron pocket at one of these momentum-space locations. The associated superconducting gap structure obtained with the generated low-energy electronic band structure from spin-fluctuation mediated pairing agrees well with that measured experimentally. Finally, from a comparison of the computed spin susceptibility to available neutron scattering data, we discuss the necessity of additional self-energy effects, and explore the role of orbital-dependent quasiparticle weights as a minimal means to include them.

\end{abstract}

\maketitle

\section{Introduction} 

The research of electronic properties of iron-based superconductors has proven to be a rich field that challenges our understanding of correlated multi-band systems. In this respect, particularly the superconducting iron-chalcogenides, FeSe and doped FeTe, have attracted considerable attention due to their unusual low-energy electronic states\cite{BoehmerKreisel_review,Coldea_2017_review,Kreisel_review}. FeSe enters an orthorhombic phase near $T_n \sim 90$ K without concomitant static magnetic order at lower temperatures, unlike most of the iron-pnictide superconductors. There is strong evidence that the rotational symmetry breaking at $T_n$ is driven by the electronic degrees of freedom, and thus constitutes a rare example of electronic nematicity\cite{Fernandes2014}.  In particular, the spectrum of low-energy spin excitations, thought to drive electron pairing, is extremely anisotropic in untwinned crystals\cite{Chen2019}. The superconducting phase sets in around $T_c \sim 9$ K and is therefore generated from an instability of the nematic  ``normal'' state. Thus, it is not surprising that the superconducting properties also break rotational symmetry, as is indeed observed experimentally\cite{Sprau2017}.
However, the origins of the extreme normal state nematicity, as well as the reasons for the absence of magnetism at ambient pressure, are still being debated\cite{BoehmerKreisel_review,Coldea_2017_review,Kreisel_review}. 

The very high level of rotational anisotropy is evident, e.g., from theoretical modelling of experimental data measured at low temperatures in FeSe\cite{Kreisel_review}. Quite generally, the anisotropy inherent to standard electronic bands in the orthorhombic state, is not enough to explain the much more  extreme
anisotropy observed experimentally\cite{Boehmer2013,Tanatar2016,Sprau2017,Kostin2018,Chen2019,Zhou_2020}. Thus, various interaction-driven feedback effects have been proposed that enhance the symmetry breaking\cite{Sprau2017,Kreisel2017,Benfatto2018,Hu2018}. In this respect, several works have argued for orbital-dependent quasiparticle weights $Z_\alpha$ and shown how this effect may reconcile several experimental probes with simple models\cite{deMedici2014,Sprau2017,Kostin2018,Kreisel2017,Bjornson_2020,Cercellier2019,Biswas2018}. The existence of orbital-dependent $Z_\alpha$ has been explored extensively within dynamical mean-field theory (DMFT) and related methods, and arise naturally in multi-orbital systems with different degrees of orbital contributions to the Fermi level electronic states\cite{Georges_rev_2013,deMedici2014,Biermann_review,Yi2017,Guterding2017,Medici_review}. For the specific case of FeSe, it was shown that $Z_{xy}$ must be strongly reduced from 1, but additionally that a large quasiparticle weight anisotropy of $Z_{xz}/Z_{yz}\sim 0.2-0.3$ was necessary in order to obtain agreement with experiments\cite{Sprau2017,Kostin2018,kreisel_2018} within this approach. The origin of this rather small ratio remains unsettled, but may be related to strong feedback effects on the electronic states close to a stripe magnetic quantum critical point. One of the important implications of these extracted values for the orbital weights $Z_\alpha$ was that the electron pocket at $Y$ should be nearly completely incoherent, and thus difficult to observe with spectroscopic probes; indeed, no conclusive observation of this pocket has been reported by either angular-resolved photoemission spectroscopy (ARPES)\cite{Watson2017,Pfau2019,Rhodes2020} or scanning tunneling microscopy (STM)\cite{Kostin2018}. 

Subsequently, however, laser ARPES studies reported significant $d_{yz}$ content on the $\Gamma$-centered hole pocket, as well as a strong $d_{xz}$ component apparently inconsistent with the extreme ``orbitally selective'' scenario\cite{Liu2018}.  Furthermore, 
very recent synchrotron ARPES measurements on detwinned FeSe have proposed that the $Y$ pocket may be lifted entirely from the Fermi surface rather than unobservable due to incoherence\cite{YiPRX2019,Huh2020}. The explanation of the unusual spectroscopic observations of this pocket is therefore important not only to obtain the  correct low-energy electronic band structure relevant for (detwinned) nematic FeSe, but also to ultimately understand  the origin of the strong electronic nematicity in FeSe.  

Of course, any theoretical approach for explaining the large electronic anisotropy of FeSe relies on having a reasonable starting point for inclusion of interactions in  the band structure. It was initially proposed, based on comparison to  ARPES data available at the time, that FeSe features nematicity in the $d_{xz}$ and $d_{yz}$ intra-orbital channels with form factors transforming as the irreducible representations (IRs) A$_{\rm 1g}$ and B$_{\rm 1g}$ of the given point group\cite{Mukherjee2015,liang_2015,Suzuki2015,Watson2016,Fanfarillo2016,Fedorov2016,Onari2016,PZhang2016}. More recently, however, several works have advocated for the relevance of (intra-orbital) nematicity involving also the $d_{xy}$ orbitals\cite{Jiang_16,Xing17,Eugenio2018,Morten_2019,christensen_2020,Rhodes_2020}.  Until recently, first principles approaches were unable to stabilize a nematic ground state without concomitant stripe magnetism.  However a recent density functional theory DFT+U exploration of the energy landscape found a lowest energy nematic state transforming according to
the  E$_{\rm u}$ irreducible representation of the D$_{\rm 4h}$ point group, containing significant inter-orbital nematic components\cite{Long2020}. This form of nematicity, which breaks also inversion symmetry, was shown to  produce Fermi surfaces containing only a single electron pocket\cite{Long2020}. Whether this approach has material-specific predictive power is still an open question\footnote{ Applying a similar approach to LiFeAs, for example, produces a dramatically distorted low-energy band structure with missing $d_{xy}$ band at the Fermi level.  R. Valenti, private communication.}, however, and in any case it remains important to identify the cause of nematic ordering using more transparent model-based methods.

These recent developments raise a number of important questions related to the low-energy band structure and the associated superconducting gap structure of FeSe. For example, what underlying physical interactions naturally produce nematic order that generate single-electron-pocket bands\cite{Long2020,Rhodes_2020}, and what are the consequences of this low-energy band structure for our broader understanding of this fascinating material? 

Here, motivated by several earlier theoretical studies of the effects of longer-range Coulomb interactions $V$ on band structure renormalizations of iron-based superconductors\cite{Jiang_16,Scherer2017Mar,Zantout_2019,Bhattacharyya_PRB2020_2}, we explore the nematic phase spontaneously generated by $V$. Ref. \onlinecite{Jiang_16} originally investigated these effects for FeSe, and additionally found a dominant intra-orbital $d$-wave bond nematic order arising from nearest neighbor Coulomb repulsion. These results are, however, inconsistent with the current ARPES description of the FeSe Fermi surface and deserve to be re-examined.  To this end, we have begun with the canonical DFT calculation of the FeSe band structure in the tetragonal phase\cite{Eschrig09}, and introduced nematic order driven by longer-range Coulomb interactions systematically. We find that, in addition to the generation of small electron and hole pockets by $V$, it generates {\it inter-orbital} nematicity between $d_{yz}$ and $d_{xy}$, and $d_{xz}$ and $d_{xy}$ states in a large region of parameter space at low temperatures. This form of nematic order is distinct from those discussed previously in the literature\cite{Jiang_16,Long2020,christensen_2020,Rhodes_2020}. The inter-orbital nematic order components hybridize the low-energy bands near $X$ and $Y$, respectively, and thereby allow, depending on their amplitudes, for the lifting of one of the electron pockets. We explore this ``$V$-scenario'' for nematicity and Fermi surface anisotropy, and demonstrate how it naturally generates Fermi surfaces containing only a single electron pocket. We next discuss the comparison of the spin excitations obtained using this renormalized band structure with the highly anisotropic spectrum reported in the experimental literature\cite{Chen2019}. Finally, we compute the resulting momentum-dependent superconducting gap structure, and discuss implications for other experimental probes of FeSe.  

\section{Model and Method}

In order to explore how the electronic structure of FeSe is affected by longer-range Coulomb interactions, specifically nearest-neighbor (NN) density interactions, we apply the following many-body Hamiltonian
\begin{align}\label{eq:ham}
\begin{split}
H = & \, H_{\rm kin} + H_{\rm soc} + H_{\rm int}
\\
= & - \sum_{i\,j} \sum_{\mu\nu} \sum_{\sigma} c^{\,\dagger}_{i\mu\sigma} \left( t^{\,\mu\nu}_{ij} + \mu_0 \delta_{ij} \delta_{\mu\nu} \right)  c^{\,\phantom{\dagger}}_{j\nu\sigma}
\\
 & + \lambda_{\rm SOC} \sum_{i} \sum_{\mu\nu} \sum_{\sigma\sigma'} c^{\dagger}_{i\mu\sigma} \, \bm{L}^{\mu\nu} \cdot \bm{S}^{\sigma\sigma'} c^{\phantom{\dagger}}_{i\nu\sigma'}
\\
 & + \frac{V}{2} \sum_{\langle i,j\rangle} \sum_{\mu\nu} \sum_{\sigma\sigma'} n^{\phantom{\dagger}}_{i\mu\sigma} n^{\phantom{\dagger}}_{j\nu\sigma'},
\end{split}
\end{align}
where $\langle i,j \rangle$ indicates the set of NN sites. Here, the operator $c^{\phantom{\dagger}}_{i\mu\sigma}$ annihilates an electron in orbital $d_{\mu}$ with spin projection $\sigma$ at lattice site $\bm{R}_i$, and $n_{i\mu\sigma}$ represents the density operator. The kinetic part of the Hamiltonian\cite{Eschrig09} $H_{\rm kin}$ contains the hopping matrix elements and the chemical potential $\mu_0$, $H_{\rm soc}$ includes the atomic spin-orbit coupling (SOC) with strength $\lambda_{\rm SOC}$, and $H_{\rm int}$ describes the repulsive NN density interaction. For all results presented in the present work, we adjust $\mu_0$ to keep a fixed electron filling of $\langle n \rangle = 6.0$. Note that for simplicity we completely discard the usual onsite Hubbard Coulomb repulsion $H_U$ in the model. As is well-known such interactions can lead to important band renormalization, magnetism, and nematicity\cite{Fernandes2014,Chubukov2016,Kreisel_review,Bhattacharyya_PRB2020_1,Gastiasoro2015}, none of which lift the $Y$ electron pocket. Here, we assume that $H_U$ merely renormalizes the DFT band structure, though we do not include such effects explicitly, and explore the nematic instability generated solely from $H_{\rm int}$ containing only NN Coulomb repulsion.

The symmetry properties of FeSe, and thereby $H$, are governed by the non-symmorphic space group $P4/nmm$, which consists of eight point group elements $\{g\,|\,(0,0,0)\}$ for $g \in$ D$_{\rm 2d}$, and $\{gI\,|\,(\nicefrac{1}{2},\nicefrac{1}{2},0)\}$ with $I$ denoting inversion. In order for the above to constitute a closed group, the product of space group elements is defined modulo integer lattice translations \cite{Cvetkovic2013Oct}. For both analytical tractability and numerical simplicity we perform our study in the 1-Fe unit cell, for which the space group $P4/nmm$ is lowered to the symmorphic subgroup D$_{\rm 2d}$. Although one thereby lacks certain symmetry elements of the original system, it is still possible to capture the violation of rotational symmetry and the emergence of nematic order. In particular, we need to focus only on the generators of the $\rm D_{2d}$ group $g = \{E,\, S_4,\,C_2,\,C'_{2},\,\sigma_d\}$, and the associated five IRs $\Gamma = \{\rm A_1,\,A_2,\,B_1,\,B_2,\,E\}$. Here, the generator $S_4$ refers to the combined operation of $C_4\sigma_{xy}$.

We incorporate the effects of NN Coulomb interactions by performing a Hartree-Fock mean-field (MF) decoupling of $H_{\rm int}$, and introduce the following homogeneous bond-order fields in wave-vector space \cite{Jiang_16, Scherer2017Mar}
\begin{align}\label{eq:bond_order_fields}
N^{\mu\nu}_{\bf{k}\sigma} = - \frac{2}{\mathcal{N}} \sum_{\bf{k}'} f_{\bf{k}-\bf{k}'}^{\rm A_1} \langle c^{\dagger}_{\bf{k}'\nu\sigma} c^{\phantom{\dagger}}_{\bf{k}'\mu\sigma} \rangle,
\end{align}
with $\mathcal{N}$ being the number of lattice sites, and $f_{\bf{k}}^{\rm A_1}$ being the form factor of the interaction, which transforms as the IR $\rm A_1$. Specifically for the nearest-neighbor interaction studied here $f_{\bf{k}}^{\rm A_1} \propto \sum_{{\rm d}} \cos (k_{\rm d}\,a_{\rm d})$, where $\rm d$ labels the set of primitive lattice vectors, and $a_{\rm d}$ is the lattice constant in the $\rm d$-direction. Note that in the decoupling of $H_{\rm int}$, the Hartree terms contribute only to a constant energy shift, which can be readily absorbed in the chemical potential $\mu_0\mapsto\mu_0-4V\sum_{\bf{k}'\mu\sigma}\langle n_{\bf{k}'\mu\sigma}\rangle/\mathcal{N}$.

\begin{figure}[t]
    \centering
    \includegraphics[width = 0.91 \columnwidth]{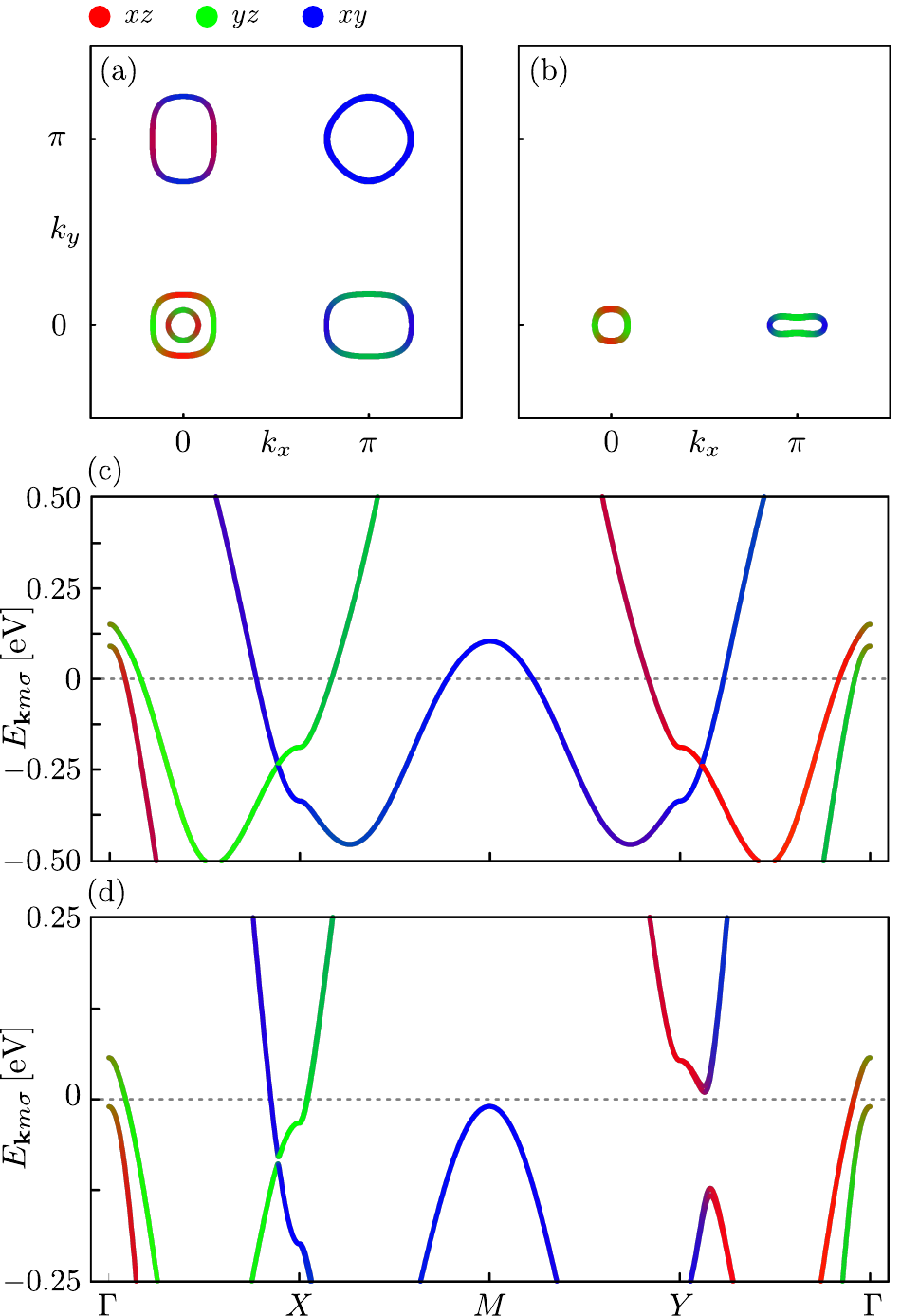}
    \caption{Orbitally resolved Fermi surfaces (FSs) and band structures in the absence (a),(c) and presence (b),(d) of NN interaction effects. The latter were obtained with $V = 0.45\,$eV and $\alpha = 2.8$. The band was adopted from Ref.~\cite{Eschrig09}, with $k_z = 0$ and SOC coupling in the spin $z$-direction, $\lambda_{\rm SOC}=30\,$meV. In (d) we clearly see the band shifts and gap openings discussed in the main text, resulting in the highly anisotropic FS displayed in panel (b). All figures were obtained with $k_{\rm B}T = 1\,$meV and $k_{\rm B}\equiv 1$.}
    \label{fig:FSs_and_band}
\end{figure}

Upon reaching the nematic transition temperature $T_n$, the system will spontaneously violate $S_4$ symmetry, and we therefore seek to split the bond order fields into $S_4$ symmetry-preserving and symmetry-breaking terms, in order to identify the origin of nematicity driven by NN interactions. The symmetry-preserving terms, denoted by $N_{{\bf k}\sigma,{\rm br}}^{\mu\nu}$,  will in general lead to band renormalizing (br) terms, and is obtained by averaging over the four cycles of $S_4$
\begin{align}\label{eq:av_s4}
N_{\bf{k}\sigma,{\rm br}}^{\mu\nu} = \frac{1}{4} \sum_{\ell=0}^3 \sum_{\mu'\nu'} \big[ D(S^{\ell}_4) \big]^{\mu\mu'} \, N^{\mu'\nu'}_{S_4^{-\ell}\bf{k}\,\sigma} \, \big[D^{\dagger}(S^{\ell}_4)\big]^{\nu'\nu},
\end{align}
where $D(S_{4}^{\ell})$ is the representation of $S_4^{\ell}$ in orbital space. As a consequence of SOC, one should in general also perform the $S_4$ averaging in spin space. However, by exploiting that the fields $N_{{\bf k}\sigma}^{\mu\nu}$ are diagonal in spin space and the fact that the action of $S_4$ is equivalent to a $\pi/2$-rotation about the spin $z$-axis, we find that $S_4$ leaves the spin projection $\sigma$ of $N_{{\bf k}\sigma}^{\mu\nu}$ invariant. 
Momentum dependent band renormalizing terms, such as $N_{{\bf k}\sigma,{\rm br}}^{\mu\nu}$, have been previously put forward as potential candidates for explaining the band structure renormalizations of iron-pnictides and iron-chalcogenides\cite{Jiang_16,Scherer2017Mar,Bhattacharyya_PRB2020_1}. 

Here, the main focus is on the $S_4$ symmetry breaking (sb) terms leading to nematic order, which are simply the remainder of the total field $N_{\bf{k}\sigma,{\rm sb}}^{\mu\nu} = N_{\bf{k}\sigma}^{\mu\nu} - N_{\bf{k}\sigma,{\rm br}}^{\mu\nu}$. In order to extract the form factors entering in the symmetry-preserving and symmetry-breaking fields, we perform the following projection onto the normalized lattice versions of the basis functions $f_{\bf{k}}^{\gamma}$ belonging to the group $\rm D_{2d}$
\begin{align}
    N_{\bf{k} \sigma,{\rm br/sb}}^{\mu\nu} = \sum_{\gamma}f^{\gamma}_{\bf{k}}N_{\gamma\sigma,{\rm br/sb}}^{\mu\nu}.
\end{align}
As a consequence of the NN interaction, the sum over $\gamma$ is restricted to include only the lowest order lattice harmonics, i.e. only linear combinations of $\cos(k_{\rm d}a_{\rm d})$ and $\sin(k_{\rm d}a_{\rm d})$ enter in $f_{\bf{k}}^{\gamma}$, while a longer-ranging interaction in general allows for higher order terms. Thus, we arrive at the following MF decoupled interaction 
\begin{align}\label{eq:ham_mf_int}
H^{\rm MF}_{\rm int} = V\sum_{{\bf{k}},\gamma}\sum_{\mu\nu}\sum_{\sigma}c^{\,\dagger}_{\bf{k}\mu\sigma}f_{\bf{k}}^{\gamma}\left( N_{\gamma\sigma,{\rm br}}^{\mu\nu} + \alpha N_{\gamma\sigma,{\rm sb}}^{\mu\nu} \right)c^{\,\phantom{\dagger}}_{\bf{k}\nu\sigma}.
\end{align}
Note that we here introduced the enhancement factor $\alpha$, since the interaction strength $V$ in the symmetry-preserving and symmetry-breaking channels in general will be different, since the fields belong to different IRs of the point group, and can potentially lead to distinct couplings in a renormalization group procedure. Thus we allow $\alpha\neq 0$ throughout, similar to earlier works\cite{Jiang_16,Scherer2017Mar}. 

Lastly, in App.~\ref{app:details_on_bond_order_fields} we elaborate further on the construction of $N_{\gamma\sigma,{\rm br/sb}}^{\mu\nu}$, display their explicit matrix structure, and list the basis functions $f_{\bf{k}}^{\gamma}$ used in the upcoming sections.

\section{Results} 

Equipped with the above model Hamiltonian and symmetry considerations, we are now able to calculate the bond order fields $N_{\gamma\sigma,{\rm br/sb}}^{\mu\nu}$ self-consistently for a fixed electron filling $\langle n \rangle = 6.0$, (for details see App.~\ref{app:details_on_bond_order_fields}). In our calculations we adopt the kinetic Hamiltonian for FeSe given in Ref.~\cite{Eschrig09}, after including SOC in the spin $z$-direction of bare strength $\lambda_{\rm SOC}=30\,$meV. However, the exact value of SOC is not qualitatively important for the emergence of inter-orbital nematic components and the concomitant disappearance of one of the electron pockets, as found further below. However, SOC is important, within the current model, for the FS to contain only a single hole pocket at $\Gamma$. Due to weak inter-layer coupling in the $c$ direction, we  restrict our study to the quasi-2D lattice of a single layer of Fe and Se atoms. The resulting orbitally-resolved Fermi surface (FS) and band structure in the absence of NN interactions are displayed in Fig.~\ref{fig:FSs_and_band}(a) and \ref{fig:FSs_and_band}(c), respectively. 

\begin{figure}
    \centering
    \includegraphics[width = \linewidth]{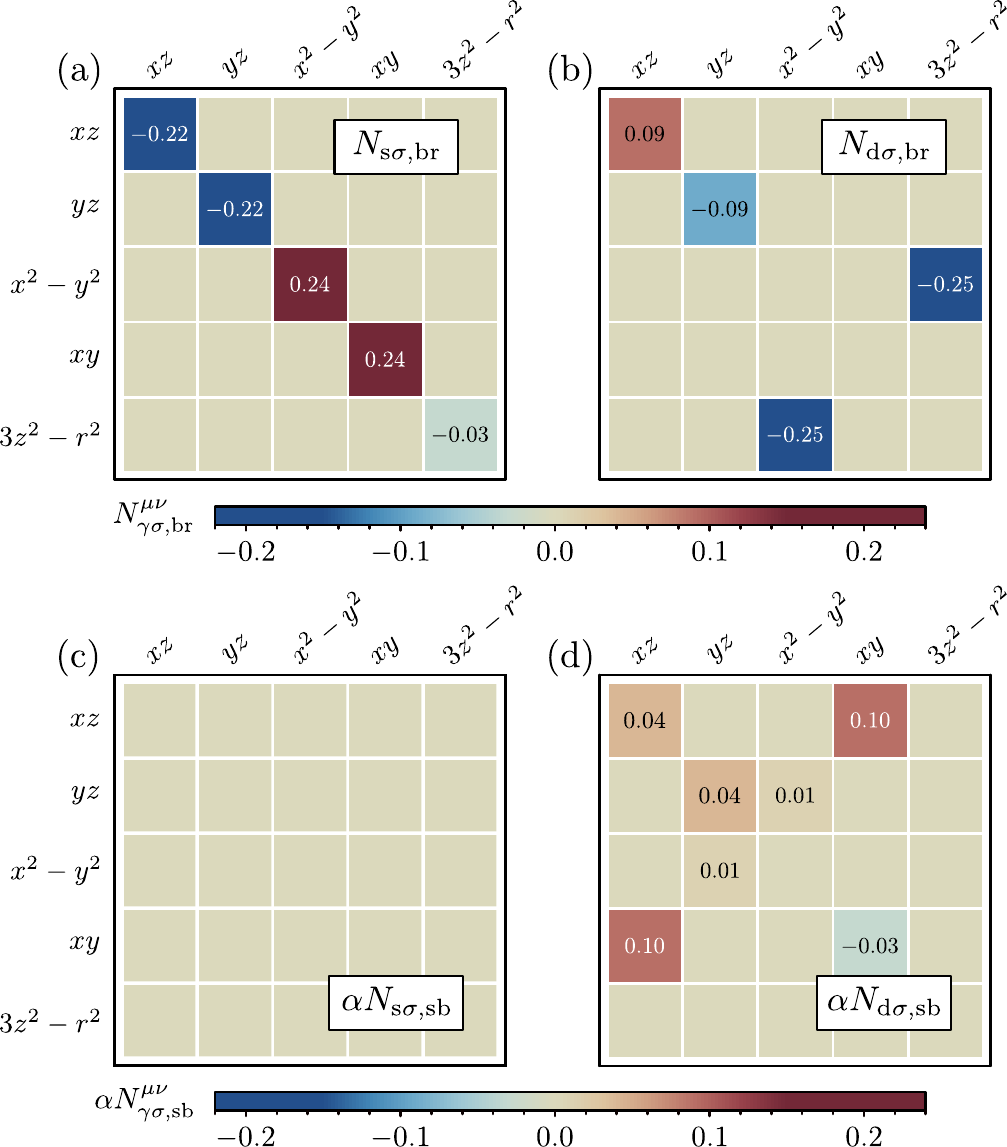}
    \caption{Band renormalizing (br) and $S_4$ symmetry breaking (sb) bond-order fields $N^{\mu\nu}_{\rm \gamma\sigma,br/sb}$, calculated self-consistently for the parameters and temperature used in Fig.~\ref{fig:FSs_and_band}(b) and (d), and fixed electron filling $\langle n \rangle = 6.0$, see App.~\ref{app:details_on_bond_order_fields} for further details. The fields in (a),(c) couple to the form factor $f_{\bf{k}}^{\,\rm s}=\cos k_x + \cos k_y$, while the ones displayed in (b),(d) couple to $f_{\bf{k}}^{\,\rm d}=\cos k_x - \cos k_y$. The values of all matrix elements have for clarity been rounded to the second decimal place. For brevity we only display $\gamma=\{\rm s,\,d\}$, but additional plots of the remaining fields can be found in App.~\ref{app:details_on_bond_order_fields}.
    }
    \label{fig:fields}
\end{figure}

The effects of NN interactions, captured by the terms entering in Eq.~\eqref{eq:ham_mf_int}, give rise to the two aforementioned effects, namely band renormalization and $S_4$ symmetry breaking. Let us momentarily discuss the former, $N_{\gamma\sigma,{\rm br}}^{\mu\nu}$, by focusing on the low-energy $t_{2g}$ orbitals. For this case, the fields lead to the following two effects: $i)$ collective down-shift of the hole pockets at $\Gamma$ and $M$, and $ii)$ up-shift of the Dirac points at $X$ and $Y$. The down-shift of the hole pockets is readily attributed to the self-consistent fields in Fig.~\ref{fig:fields}(a), which couple to the form factor $f_{\bf{k}}^{\rm \,s}=\cos k_x + \cos k_y$. Specifically the fields $N_{{\rm s}\sigma,{\rm br}}^{xz\,xz}=N_{{\rm s}\sigma,{\rm br}}^{yz\,yz}=-0.22$ lead to a down-shift of $d_{xz}$,$d_{yz}$-dominated bands at $\Gamma$, while $N^{xy\,xy}_{{\rm s}\sigma,{\rm br}}=0.24$ ensures the down-shift of $d_{xy}$-dominated bands at $M$. The latter occurs since the form factor has a relative sign on the two pockets $f_{\Gamma}^{\,\rm s}=-f_{M}^{\,\rm s}$. Additionally we note that the effects of the $\rm s$-wave fields on the electron pockets are minimal since $f^{\,\rm s}_{X}=f^{\,\rm s}_{Y}=0$.

By contrast to the above, the fields presented in Fig.~\ref{fig:fields}(b) are dominant at the $X$ and $Y$ points due to the $\rm d$-wave form factor $f_{\bf{k}}^{\, \rm d}=\cos k_x - \cos k_y$. In fact, the field $N_{{\rm d}\sigma,{\rm br}}^{xz\,xz}=0.09$ ($N_{{\rm d}\sigma,{\rm br}}^{yz\,yz}=-0.09$) shifts $d_{xz}$($d_{yz}$)-dominated bands at $Y$ ($X$), pushing the Dirac points up closer to the Fermi level. Similar to the $\rm s$-wave fields, also here the symmetry of the form factor, $f^{\,\rm d}_{X}=-f^{\,\rm d}_{Y}$, compensates the relative sign between the two fields $N^{xz\,xz}_{\rm d\sigma,br}/N^{yz\,yz}_{\rm d\sigma,br}=-1$. Furthermore, we point out that the electron pockets acquire a peanut-like shape for appropriate values of $V$, since the fields $N_{\rm d\sigma,br}^{\mu\nu}$ do not affect the $d_{xy}$ orbitals. Lastly note that the down- (up-) shift of the hole pockets (Dirac points) leads to smaller $\Gamma$ and $M$ ($X$ and $Y$) pockets. In fact, it was previously shown that $N_{\bf{k}\sigma,{\rm br}}^{\mu\nu}$ can completely remove the $M$ pocket in FeSe, while for LiFeAs it was the $\Gamma$ pockets which were pushed away from the Fermi level \cite{Bhattacharyya_PRB2020_1}. 

Turning our attention to the symmetry breaking fields, $N_{\gamma\sigma,{\rm sb}}^{\mu\nu}$, we find that only fields that are coupled to the $\rm d$-wave form factor play an essential role, see Fig.~\ref{fig:fields}(c) and \ref{fig:fields}(d). By applying the same logic as for the band renormalizing terms, we observe that the field $N_{\rm d\sigma,sb}^{xz\,xz}$ ($N_{\rm d\sigma,sb}^{yz\,yz}$) leads to an up- (down-) shift of $d_{xz}$($d_{yz}$)-dominated bands at $Y$ ($X$). While this effect appears similar to the one encountered for $N_{\rm d\sigma,br}^{\mu\nu}$, we stress that here the Dirac points are shifted in opposite directions due to the violation of $S_4$-symmetry. This reduction in symmetry also allows for anisotropic inter-orbital $d_{xz}-d_{xy}$ and $d_{yz}-d_{xy}$ hybridization terms, namely $\alpha N^{xz\,xy}_{\rm d\sigma,sb}\approx 0.10$ and a vanishing coupling $\alpha N^{yz\,xy}_{\rm d\sigma,sb}$. It is in fact these particular couplings which lead to the distinct hybridization gaps at $X$ and $Y$ evident from Fig.~\ref{fig:FSs_and_band}(d), and they will, in synergy with all the effects discussed above, result in the highly anisotropic FS shown in Fig.~\ref{fig:FSs_and_band}(b), featuring only a single electron pocket at the $X$ point. Thus, as a function of temperature, a Lifshitz transition necessarily takes place such that a shrinking $Y$-pocket eventually disappears at a certain temperature below $T_n$. A detailed discussion of the experimental evidence for such a temperature-induced Lifshitz transition can be found in Ref.~\onlinecite{Rhodes_2020}.

 To gain deeper insight into the emergent nematic order, we now proceed and study more carefully the IRs of $N_{{\bf k}\sigma,\rm br/sb}^{\mu\nu}$. By construction, the $S_4$ symmetry preserving fields can only consist of terms transforming as A$_1$ and A$_2$, and, in fact, we find through our self-consistent calculations that only A$_1$ terms are non-zero in $N_{{\bf k}\sigma,{\rm br}}^{\mu\nu}$, see App.~\ref{app:ir_fields} for further details.

\begin{figure}[b]
    \centering
    \includegraphics[width = \columnwidth]{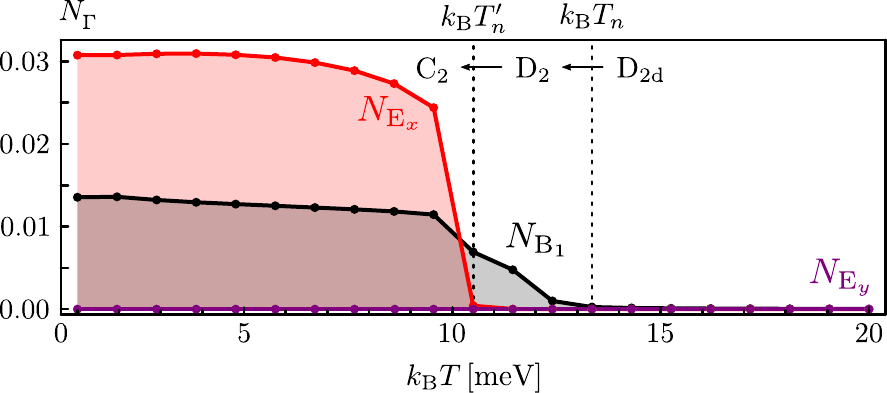}
    \caption{Leading order parameters $N_{\rm B_1}$ and $N_{\rm E}$ versus temperature. The system becomes nematic at $T_n$, and displays a second phase transition at $T_n'$. For the latter, $N_{{\rm E}_x}$ becomes non-zero, allowing for hybridization terms at $Y$. In this figure, we used the same parameters as in Figs.~\ref{fig:FSs_and_band} and \ref{fig:fields}.}
    \label{fig:temp_sweep}
\end{figure}

In contrast, $N_{{\bf k}\sigma,{\rm sb}}^{\mu\nu}$ can contain terms transforming as the remaining three IRs of the group, i.e. $\rm B_1$, $\rm B_2$ and $\rm E$, allowing for the nematic order to arise in any of these three channels. As we explicitly show in App.~\ref{app:ir_fields}, we find non-zero $S_4$ symmetry breaking terms belonging to two distinct IRs, namely B$_1$ and E. This implies that the system undergoes two consecutive phase transitions upon lowering of the temperature. In order to quantify this, we focus on the $t_{2g}$ orbitals, and define the following leading order parameters
\begin{align}
\begin{split}
    N_{\rm B_1} &= (N_{\rm d\sigma, sb}^{xz\,xz} + N_{\rm d\bar{\sigma}, sb}^{xz\,xz} + N_{\rm d\sigma,sb}^{yz\,yz} + N_{\rm d\bar{\sigma},sb}^{yz\,yz}) / 4,
    \\
    N_{\rm E} &= \left([N_{{\rm d}\sigma,{\rm sb}}^{xz\,xy} + N_{{\rm d}\bar{\sigma},{\rm sb}}^{xz\,xy}] / 2, [N_{{\rm d}\sigma,{\rm sb}}^{yz\,xy} + N_{{\rm d}\bar{\sigma},{\rm sb}}^{yz\,xy}] / 2\right)
    \\
    &\equiv (N_{{\rm E}_x}, N_{{\rm E}_y}),
    \end{split}
\end{align}
where $\bar{\sigma}$ is the opposite spin projection of $\sigma$. For details see App.~\ref{app:ir_fields}. In Fig.~\ref{fig:temp_sweep} we show the values of these order parameters for various temperatures, and indeed find that $N_{\rm B_1}$ becomes non-zero at $T_n$, while only the single component $N_{{\rm E}_{x}}$ condenses at lower temperatures $T'_n$. An intra-orbital nematic order thus arises at $T_n$, lowering the point group symmetry from $\rm D_{\rm 2d}$ to $\rm D_{2}$, and leads to the up- (down-) shift of $d_{xz}$($d_{yz}$)-dominated bands at $Y$ ($X$) discussed in the previous paragraph. The second transition at $T_n'$ further reduces the point group to $\rm C_2$ and allows for anisotropic inter-orbital $d_{xz}-d_{xy}$ or $d_{yz}-d_{xy}$ hybridization terms, i.e. the effects ultimately leading to the highly anisotropic FS, shown in Fig.~\ref{fig:FSs_and_band}(b).

\begin{figure*}[t]
    \includegraphics[width = \linewidth]{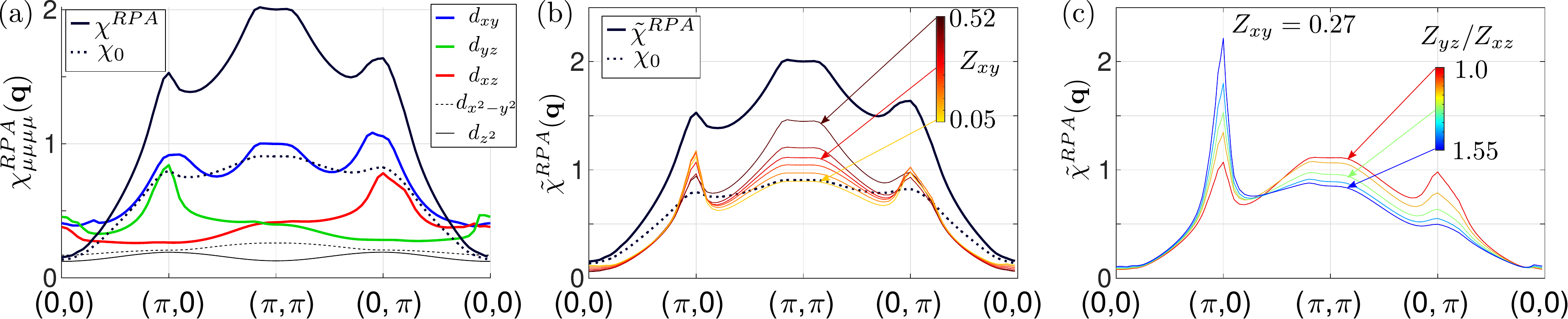}
    \caption{(a) Spin susceptibility as calculated from our model including the nematic order induced by nearest-neighbor Coulomb interactions, but using a fully coherent electronic structure, $Z_\alpha=1$, $U=0.805\,$eV, $J/U=1/6$. (b) Spin susceptibility as obtained from a weakly correlated system by assuming orbitally selective quasiparticle weights\cite{Bjornson_2020} (dark red curve) ($U=3.1\,$eV) and successively decreasing the quasiparticle weight for the $d_{xy}$ orbital to almost negligible quasiparticle contribution (yellow, $U=3.65\,$eV). (c) Moderate correlation with a small $Z_{xy}=0.27$, but additionally splitting the quasiparticle weights between the $d_{xz}$ and $d_{yz}$ orbitals yields a strongly anisotropic susceptibility with no visible peak at $(0,\pi)$ starting from $Z_{yz}/Z_{xz}\approx 1.3$ ($U$ decreased slightly to not cross the magnetic instability).}\label{fig:chi}
\end{figure*}

We stress that the effects of the $e_g$ orbitals and remaining fields $N^{\mu\nu}_{\gamma\sigma,{\rm br/sb}}$ not displayed in Fig.~\ref{fig:fields}, are all incorporated in our calculations, but do not lead to qualitative changes of the low-energy band structure and FS, and are therefore not explicitly mentioned in the above discussion. For a complete overview of all fields  $N_{\gamma\sigma,{\rm br/sb}}^{\mu\nu}$, see App.~\ref{app:details_on_bond_order_fields}. Furthermore, we note that the final nematic band structure, as seen e.g. in Fig.~\ref{fig:FSs_and_band}(d), obviously depends on the starting point, i.e. the tetragonal DFT band structure. Thus, the final quantitative energy scales, i.e. the hybridization gap at $Y$ and the required amplitudes of $V$ and $\alpha$ for generating a FS similar to that shown in Fig.~\ref{fig:FSs_and_band}(b), depend on the initial bare band structure. Lastly, we note that the purely intra-orbital nematic order generated from NN Coulomb repulsion found earlier~\cite{Jiang_16,Scherer2017Mar}, can be reproduced here when applying the same band structure as in Ref.~\onlinecite{Jiang_16}. We have not located the exact band property that leads to the additional inter-orbital nematic fields from the DFT bulk FeSe band model used in the current study\cite{Eschrig09}, but note that the solution presented in Fig.~\ref{fig:FSs_and_band} is quite generic at low temperatures and exists for a wide parameter range.

The low-energy electronic structure established above has important consequences for spin excitations that should be compared with experiments. In this respect, a series of inelastic  neutron scattering experiments\cite{Rahn2015,Wang2016A,Chen2019} have established a remarkable set of magnetic phenomena in FeSe.  At low temperatures in the nematic phase, but high energies $\sim$ 100 meV, strong  $(\pi,\pi)$ (N\'eel)  fluctuations dominate the spectrum, with weaker but still prominent $(\pi,0)$ and $(0,\pi)$ (stripe-like) fluctuations.  As the energy is lowered to a few tens of meV, a spin gap develops in the  $(\pi,\pi)$ spectrum, but $(\pi,0)$ spin fluctuations strengthen.  Notably, the only measurement of {\it detwinned } FeSe crystal finds that at low energies the intensity of the $(0,\pi)$ fluctuations essentially vanishes\cite{Chen2019}.  This extraordinary result should emerge from a proper theory of low energy spin and orbital degrees of freedom.  We show now that, counter-intuitively, the current proposal for  low-energy nematic electronic structure is not sufficient to explain the above-mentioned magnetic properties, and requires the additional physics of orbital selective correlations.  

In Fig. \ref{fig:chi}(a), we first illustrate the bare magnetic susceptibility Re$\chi_0({\bf q},\omega=0)$,  together with the enhanced susceptibility obtained in the random phase approximation $\chi^{RPA}$.  The spectrum is clearly dominated by intense $(\pi,\pi)$ fluctuations arising from scattering between the $d_{xy}$ states, which are not observed in experiment.    Furthermore, $(\pi,0)$ and $(0,\pi)$ states are nearly degenerate, in contradiction to the results of Ref. \onlinecite{Chen2019}.  Note that these issues are also common to conventional spin-fluctuation theories of FeSe including a $Y$ pocket\cite{kreisel_2018,Fanfarillo_mismatch2018}, or other novel schemes to lift the $Y$ pocket\cite{Rhodes_2020} via nematic order.  In Ref. \onlinecite{kreisel_2018}, it was proposed that they could be resolved by assuming orbitally selective incoherence of $d_{xy}$, $d_{xz}$, and $d_{yz}$ states, such as should take place according to previous theory\cite{Medici_review} and as observed in some experiments\cite{Yi2017}.  With a phenomenological insertion of orbitally dependent quasiparticle weights $Z_\alpha$, with $\alpha$ an orbital index, Kreisel {\it et al.}\cite{kreisel_2018} could fit inelastic neutron data on FeSe using very strong suppression of $d_{xy}$ weights and somewhat smaller suppression of $d_{xz}$ and $d_{yz}$ weights, assuming also a large ratio of at least 1.7 for the ratio $Z_{yz}/Z_{xz}$. 

We therefore explore how orbital incoherence could improve the agreement of the susceptibility calculated from the current highly nematic electronic structure with experiment.   In Fig. \ref{fig:chi}(b),  we  present the results for weak correlations, and  the evolution of the susceptibility as  correlations are enhanced by a substantial suppression of the $d_{xy}$ quasiparticle weight (see Appendix \ref{app:susc} for a brief discussion of the Ansatz of Kreisel {\it et al.}\cite{kreisel_2018}).   As anticipated from the contribution of the orbitally resolved susceptibility of the $d_{xy}$ orbital (see Fig. \ref{fig:chi}(a)), with reduction of $Z_{xy}$ comes the 
suppression of the N\'eel peak and concomitant moderate enhancement of the $(\pi,0)$ stripe peak.  On the other hand the $(\pi,0)/(0,\pi)$ anisotropy is still much weaker than that reported in Ref.  \onlinecite{Chen2019}.  In Fig. \ref{fig:chi}(c), we therefore show the effect of additionally increasing the $Z_{yz}/Z_{xz}$ anisotropy.  It is easy to see that much larger $(\pi,0)/(0,\pi)$ anisotropies are obtained in this case, but also that substantially smaller quasiparticle weight ratios, of order $\sim 1.3$, are required compared to Ref. \onlinecite{kreisel_2018}, due to the effect of inter-orbital nematic order introduced here.

\begin{figure*}[tb]
    \includegraphics[width = \linewidth]{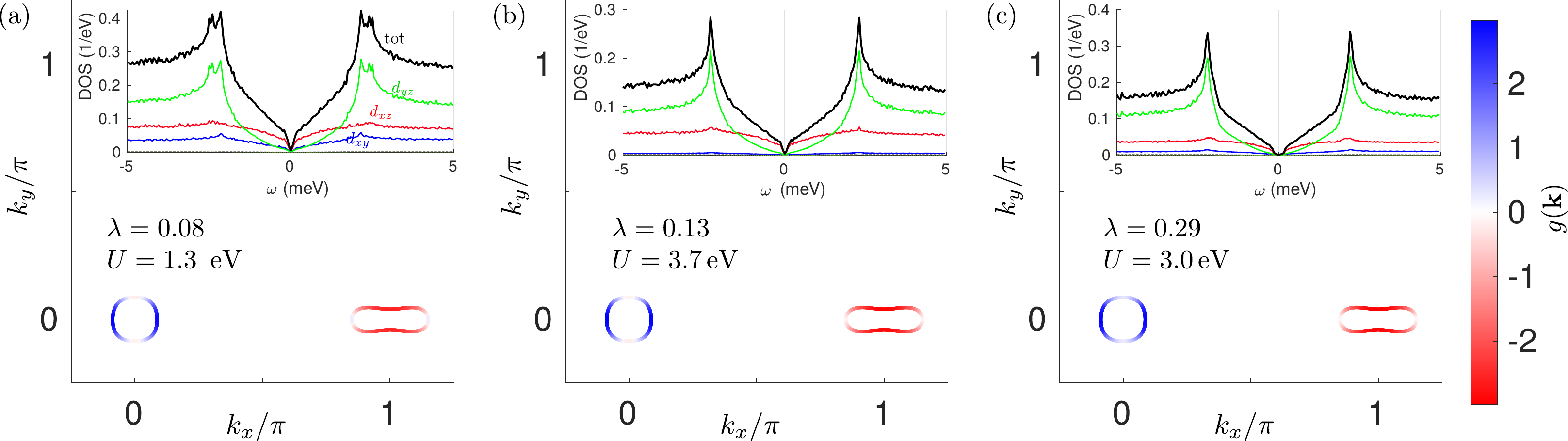}
    \caption{(a) Superconducting gap symmetry function $g(\bf{k})$ as calculated from a fully coherent electronic structure showing the strongly anisotropic gap on the two pockets. The expected spectrum shows nodal features and the eigenvalue $\lambda$ is sizeable but small. (b) $g(\bf{k})$ calculated from a electronic structure with very incoherent $d_{xy}$ orbital; almost no effect is visible except that the eigenvalue can be larger since the relative contribution of the $(\pi,0)$ scattering is higher. (c) Same quantity, but calculated including a moderately correlated $d_{xy}$ orbital and a nematic splitting of the quasiparticle weights in the $d_{yz}$ and $d_{xz}$ orbitals, yielding an order parameter with a tiny true gap, in contrast to nodes in (a) and (b).}\label{fig_gap}
\end{figure*}

On the other hand, these various hypothetical renormalizations do not lead to substantial changes in the gap structure obtained for FeSe within the corresponding spin fluctuation pairing theory.  This is simply because in a multiband system even at $\omega=0$ contributions to $\chi(\bf q)$ arise from states tens or even hundreds of meV from the Fermi level\cite{Kreisel15}.  By contrast, Fermi surface states determine the anisotropy of the effective pairing interaction completely.  This effect can be seen easily in Fig. \ref{fig_gap}, where we plot the leading eigenvector of the linearized gap equation (see Appendix \ref{app:susc}) for the same three cases described in Fig. \ref{fig:chi}.  The overall structure of the gaps and the density of states are seen to be virtually identical.

\section{Discussion and conclusions} 

In this work we have pursued a scenario where nematicity originates entirely from NN Coulomb repulsion, even though it is well-known that onsite interactions alone may also drive a nematic instability as a precursor to stripe magnetism\cite{Fernandes2014}. Interestingly, orbitally resolved studies within the spin-nematic onsite interaction-only scenario, do find sizable inter-orbital nematic susceptibilities\cite{Morten_2016}. Such studies, however, have only been performed in the tetragonal paramagnetic phase, and it remains to be seen whether spontaneous breaking in the inter-orbital channel, similar to the current proposal, can also arise from higher order processes solely from onsite interactions. However, the absence of magnetism in FeSe and the overall agreement of the here-presented results to the experimental facts, raises the question of whether indeed NN-repulsion is the generator of nematicity for FeSe. From cRPA\cite{Miyake_2010} and DMFT\cite{Yin2011} calculations it is known that interactions are generally larger in FeSe than any of the other iron-based superconducting systems, presumably because a lack of intervening spacer layers reduces the screening.  Onsite interactions $U$, $J$ are known to lead to band renormalizations, in particular Fermi surface pocket shrinkage\cite{Zantout_2019,Bhattacharyya_PRB2020_2}  and orbital-dependent band narrowing. It is tempting to speculate that this, in turn, effectively enhances the importance of the unusually large $V$ in FeSe, thereby boosting instabilities driven by this channel.

Recently, another theoretical study investigated the possibility of nematic order lifting one of the electron pockets (above the Fermi level) in FeSe\cite{Rhodes_2020}. In agreement with the present work, a Lifshitz transition necessarily takes place as a function of temperature, and a superconducting gap structure consistent with experiments follows directly from the resulting FS with the missing $Y$-pocket. A main differences between Ref.~\onlinecite{Rhodes_2020} and the current approach is the nature of the nematic order; whereas Rhodes {\it et al.}\cite{Rhodes_2020} begin with a phenomenological $\bf k\cdot\bf p$ expansion around $\Gamma$, $X$ and $Y$, and explore the role of a large  intra-orbital B$_{\rm 1g}$ $d_{xy}$-nematic order parameter imposed by ``hand'' on the band structure, together with   additional onsite Hartree shifts to this orbital, we have started from a nearest-neighbor Coulomb interaction, and shown that (self-consistently generated) inter-orbital $d_{xy}-d_{xz/yz}$ nematic components lift the $Y$-pocket. Thus, while a B$_{\rm 1g}$ intra-orbital $d_{xy}$ nematic component is also present in our approach as seen from Fig.~\ref{fig:fields}(d), the most important nematic components for generating a FS containing no electron $Y$-pocket are the distinct inter-orbital components.  

 As mentioned in the ``Results'' section, the nematic lowest-temperature phase advocated in this work, resides in the $\rm C_2$ (monoclinic) group, containing only a $C_2$ rotation around the $x$-axis as the remaining symmetry operation. This constitutes a clear prediction within the current scenario; the electronic sector should exhibit a double transition as the temperature is lowered. We are not aware of such evidence, e.g. from specific heat data\cite{Jiao_2017,MURATOV2018785,Boehmer2015}, which may be because of the very small electronic entropy change at the second transition, or simply because the two transitions are accidentally close (in temperature) (see App. \ref{app:ir_fields} for more details) in FeSe. In addition, if the nematic order couples strongly enough to the atomic lattice, such symmetry lowering could be tested by experiments sensitive to the overall crystal point group. In this regard, however, we note that recent  experimental and theoretical studies have advocated for a more complex static crystal microstructure in FeSe than previously thought. Locally, FeSe appears to accommodate a myriad of inhomogeneous nanoscale lattice distortions, where only the spatially averaged structure complies with the standard tetragonal (orthorhombic) crystal symmetries at high (low) temperatures\cite{Koch_2019,Frandsen_2019,Long2020,Zunger_2020}. This may in fact be the result of two nematic channels very close in in energy. Lastly, it is tempting to speculate that the inter-orbital nematic order discussed here, may be relevant for the recent experimental studies of structural transitions in Bi$_{1-x}$Sr$_x$Ni$_2$As$_2$, which exhibit transitions from a high-temperature tetragonal phase to a triclinic low-temperature  phase\cite{Eckberg2020}.

Recent theoretical works explored the possibility of pinned local nematic order\cite{Gastiasoro2014,Wang_Gastiasoro_2015,Steffensen_2019}. In particular, Ref.~\onlinecite{Steffensen_2019} explored the local disorder-induced nematicity from nonmagnetic impurities in the tetragonal phase at $T>T_n$. Several experimental works have reported evidence for such local nematic order above $T_n$\cite{Koch_2019,Frandsen_2019}. The results of Ref.~\onlinecite{Steffensen_2019} were obtained by applying a one-band model and relied on interactions leading to a single-component $\rm B_{\rm 1g}$ nematic order, and hence the question arises how local pinned nematic order gets affected by the presence of the substantial inter-orbital components found in this work? We have answered this question by performing a real-space calculation similar to that of Ref.~\onlinecite{Steffensen_2019}, but generalized to the multi-orbital case. For the intra-orbital $\rm B_1$ order parameter $N_{\rm B_1}$ studied here, we observe that it enters the Landau free energy expansion in the exact same manner as the order parameter studied in Ref.~\onlinecite{Steffensen_2019}. Therefore the local impurity-induced order exhibits a "flower-shape" pinned by nonmagnetic impurities\cite{Steffensen_2019}. However, for the lower temperature phase, the spatially dependent local nematic impurity-induced order is distinct from that of Ref.~\onlinecite{Steffensen_2019}, since the order parameter $N_E$ couples differently to the nonmagnetic impurity. It remains an interesting future study to investigate the detailed experimental consequences of these various local nematic orders.

In summary, we have discovered an inter-orbital nematic order generated from nearest-neighbor Coulomb repulsion for electronic models relevant for FeSe in particular, and perhaps the iron-based systems more broadly. A natural property of this kind of nematic order is the generation of highly anisotropic Fermi surfaces, featuring in some cases, only a single hole and electron pocket. We have shown how, for FeSe, this explains photoemission data and the experimentally extracted very anisotropic superconducting gap structure. However, a consistent picture of the neutron scattering data, cannot be straightforwardly obtained within this nematic scenario, without invoking additional self-energy effects, which in the simplest case, involves sizable corrections in the form of reduced orbitally dependent quasiparticle weights.

\section*{Acknowledgements} We thank M. H. Christensen and I. Eremin for insightful conversations. D.S. and B.M.A. acknowledge support from the Carlsberg Foundation. B. M. A. acknowledges support from the Independent Research Fund Denmark grant number 8021-00047B. P. J. H. was supported by the U.S. Department of Energy under Grant No. DE-FG02-05ER46236. 

\appendix
\section{Details on bond-order fields and self-consistent calculations}
\label{app:details_on_bond_order_fields}
The explicit forms of $N_{\bf{k}\sigma,{\rm br}}^{\mu\nu}$ and $N_{\bf{k}\sigma,{\rm sb}}^{\mu\nu}$ are found through the averaging in Eq.~\eqref{eq:av_s4}, and the difference $N_{\bf{k}\sigma,{\rm sb}}^{\mu\nu} = N_{\bf{k}\sigma}^{\mu\nu} - N_{\bf{k}\sigma,{\rm br}}^{\mu\nu}$, respectively. Yet this procedure can be further simplified, by relying on the projection of the fields onto the normalized lattice versions of the basis functions $f_{\bf{k}}^{\gamma}$
\begin{align}\label{eq:op_projection}
N_{\bf{k}\sigma}^{\mu\nu} = \sum_{\gamma}f_{\bf{k}}^{\gamma}\,N_{\gamma\sigma}^{\mu\nu}.
\end{align}
This projection disentangles the wave-vector and orbital dependence, and thereby greatly simplifies Eq.~\eqref{eq:av_s4}. For NN interactions in 2D systems, which are left invariant under the elements of $\rm D_{2d}$, the set of basis functions entering in Eq.~\eqref{eq:op_projection} are\newline\vspace{-1cm}
\begin{subequations}\label{eq:form_factor}
\begin{align}
    &f_{\bf{k}}^{\,\rm s} = \cos k_x + \cos k_y,  &   &f_{\bf{k}}^{\,\rm d} = \cos k_x - \cos k_y,  \\
    &f_{\bf{k}}^{{\,\rm p}_x} = \sqrt{2}\,i\sin k_x,    &   &f_{\bf{k}}^{\,{\rm p}_y} = \sqrt{2}\,i\sin k_y,
\end{align}
\end{subequations}
with $a_x=a_y=a\equiv1$. For NN interactions in three dimensions, the above equations are accompanied by $f^{\,\rm s}_{k_z}=\sqrt{2}\cos(k_zc)$ and $f^{\,\rm d}_{k_z}=\sqrt{2}\,i\sin(k_zc)$, with $a_z\equiv c \neq a$. Each basis function transforms according to one of the IRs $\Gamma$ of the point group. Specifically, $f^{\, \rm s}_{\bf k}$ ($f^{\, \rm d}_{\bf k}$) transforms as $\rm A_1$ ($\rm B_1$), while $(f_{\bf k}^{\,{\rm p}_x}, f_{\bf k}^{\,{\rm p}_y})$ transform jointly as the 2D IR E.

Next step in determining $N^{\mu\nu}_{{\bf k}\sigma,{\rm br/sb}}$, is to represent $S_4$ in spin space and in the relevant orbital basis $\{d_{xz},\,d_{yz},\, d_{x^2-y^2},\,d_{xy},\,d_{3z^2-r^2}\}$. The former is needed since the SOC in Eq.~\eqref{eq:ham} breaks spin-rotation symmetry. Additionally, we also need to determine how $S_4^{-1}$ acts on the wave-vectors. Straightforwardly, we find
\begin{align}
\begin{split}
    \mathcal{D}(S_4) &= D(S_4)\otimes D_{\rm spin}(S_4)\\ 
    &=\begin{pmatrix} 
    0  &   1  &   0 &   0   &   0\\
    - 1  &   0  &   0   &   0   &   0\\
    0   &   0   &   -1   &   0   &   0\\
    0   &   0   &   0   &   - 1 &   0\\
    0   &   0   &   0   &   0   &   1
    \end{pmatrix}\otimes \frac{\mathds{1}_{\sigma}-i\sigma_z}{\sqrt{2}},  
    \\
    S^{-1}_{4}\bf{k} &= (k_y,-k_x,-k_z),
    \end{split}
\end{align}
where $D(S_4)$ and $D_{\rm spin}(S_4)$ are matrix representations in orbital and spin space, respectively, while $\mathcal{D}(S_4)$ is the representation in the combined space. Any symmetry operation acting in spin space can be expressed in terms of the Pauli matrices $\sigma_{x,y,z}$ accompanied by the identity $\mathds{1}_{\sigma}$. Note, however, that the fields $N_{{\bf k}\sigma}^{\mu\nu}$ are diagonal in spin space, and therefore not affected by $D_{\rm spin}(S_4)$. One can therefore solely consider the action of $D(S_4)$, as discussed in connection to Eq.~\eqref{eq:av_s4}. In general, when considering a given symmetry element $g$ of the point group $\rm D_{\rm 2d}$, one needs to apply the full representation $\mathcal{D}(g)$.

By executing the above on a 2D system, we end up with the following band renormalizing (br) terms
\begin{widetext}
\begin{subequations}\label{eq:app_field_br}
\begin{align}
    N_{\rm s\sigma, br} &=
    \begin{pmatrix}
    \frac{1}{2}(N^{11}_{{\rm s}\sigma} + N^{22}_{{\rm s}\sigma})    &   \frac{1}{2}(N^{12}_{{\rm s}\sigma} - N^{21}_{{\rm s}\sigma})   &   0   &   0   &   0\\[0.2 cm]
    \frac{1}{2}(N^{21}_{{\rm s}\sigma} - N^{12}_{{\rm s}\sigma})   &   \frac{1}{2}(N^{11}_{{\rm s}\sigma} + N^{22}_{{\rm s}\sigma})     &   0  &   0   &   0\\[0.2 cm]
    0   &   0   &   N^{33}_{{\rm s}\sigma}    &   N^{34}_{{\rm s}\sigma}   &   0\\[0.2 cm]
    0   &   0   &   N^{43}_{{\rm s}\sigma}   &   N^{44}_{{\rm s}\sigma}   &   0\\[0.2 cm]
    0   &   0   &   0   &   0   &   N^{55}_{{\rm s}\sigma}    \\[0.2 cm]
    \end{pmatrix},
\end{align}
\begin{align}
    N_{\rm d\sigma,{\rm br}} &= 
    \begin{pmatrix}
    \frac{1}{2}(N^{11}_{{\rm d}\sigma} - N^{22}_{{\rm d}\sigma})    &   \frac{1}{2}(N^{12}_{{\rm d}\sigma} + N^{21}_{{\rm d}\sigma})   &   0   &   0   &   0\\[0.2 cm]
    \frac{1}{2}(N^{12}_{{\rm d}\sigma} + N^{21}_{{\rm d}\sigma})   &   -\frac{1}{2}(N^{11}_{{\rm d}\sigma} - N^{22}_{{\rm d}\sigma})     &   0 &   0   &   0\\[0.2 cm]
    0   &   0   &   0   &   0   &   N_{{\rm d}\sigma}^{35}\\[0.2 cm]
    0   &   0   &   0   &   0   &   N_{{\rm d}\sigma}^{45}\\[0.2 cm]
    0   &   0   &   N^{53}_{{\rm d}\sigma}    &   N_{{\rm d}\sigma}^{54}   &   0
    \end{pmatrix},
\end{align}
\begin{align}    N_{{\rm p}_x\sigma, {\rm br}} &= 
    \begin{pmatrix}
    0    &   0   &   \frac{1}{2}(N_{{\rm p}_x\sigma}^{13} + N_{{\rm p}_y\sigma}^{23})   &   \frac{1}{2}(N_{{\rm p}_x\sigma}^{14} + N_{{\rm p}_y\sigma}^{24})    &   \frac{1}{2}(N_{{\rm p}_x\sigma}^{15} - N_{{\rm p}_y\sigma}^{25})\\[0.2 cm]
    0   &   0     &   \frac{1}{2}(N_{{\rm p}_x\sigma}^{23} - N_{{\rm p}_y\sigma}^{13})  &   \frac{1}{2}(N_{{\rm p}_x\sigma}^{24} - N_{{\rm p}_y\sigma}^{14})   &   \frac{1}{2}(N_{{\rm p}_x\sigma}^{25} + N_{{\rm p}_y\sigma}^{15})\\[0.2 cm]
    \frac{1}{2}(N_{{\rm p}_x\sigma}^{31} + N_{{\rm p}_y\sigma}^{32})   &   \frac{1}{2}(N_{{\rm p}_x\sigma}^{32} - N_{{\rm p}_y\sigma}^{31})   &   0 &   0   &   0\\[0.2 cm]
    \frac{1}{2}(N_{{\rm p}_x\sigma}^{41} + N_{{\rm p}_y\sigma}^{42})   &   \frac{1}{2}(N_{{\rm p}_x\sigma}^{42} - N_{{\rm p}_y\sigma}^{41})   &   0 &   0   &   0\\[0.2 cm]
    \frac{1}{2}(N_{{\rm p}_x\sigma}^{51} - N_{{\rm p}_y\sigma}^{52})   &   \frac{1}{2}(N_{{\rm p}_x\sigma}^{52} + N_{{\rm p}_y\sigma}^{51})   &   0 &   0   &   0
    \end{pmatrix},
\end{align}
\begin{align}
    N_{{\rm p}_y\sigma, {\rm br}} &= 
     \begin{pmatrix}
    0    &   0   &   -\frac{1}{2}(N_{{\rm p}_x\sigma}^{23} - N_{{\rm p}_y\sigma}^{13})   &   - \frac{1}{2}(N_{{\rm p}_x\sigma}^{24} - N_{{\rm p}_y\sigma}^{14})    &   \frac{1}{2}(N_{{\rm p}_x\sigma}^{25} + N_{{\rm p}_y\sigma}^{15})\\[0.2 cm]
    0   &   0     &   \frac{1}{2}(N_{{\rm p}_x\sigma}^{13} + N_{{\rm p}_y\sigma}^{23})  &   \frac{1}{2}(N_{{\rm p}_x\sigma}^{14} + N_{{\rm p}_y\sigma}^{24})   &   - \frac{1}{2}(N_{{\rm p}_x\sigma}^{15} - N_{{\rm p}_y\sigma}^{25})\\[0.2 cm]
    - \frac{1}{2}(N_{{\rm p}_x\sigma}^{32} - N_{{\rm p}_y\sigma}^{31})   &   \frac{1}{2}(N_{{\rm p}_x\sigma}^{31} + N_{{\rm p}_y\sigma}^{32})   &   0 &   0   &   0\\[0.2 cm]
    - \frac{1}{2}(N_{{\rm p}_x\sigma}^{42} - N_{{\rm p}_y\sigma}^{41})   &   \frac{1}{2}(N_{{\rm p}_x\sigma}^{41} + N_{{\rm p}_y\sigma}^{42})   &   0 &   0   &   0\\[0.2 cm]
    \frac{1}{2}(N_{{\rm p}_x\sigma}^{52} + N_{{\rm p}_y\sigma}^{51})   &   - \frac{1}{2}(N_{{\rm p}_x\sigma}^{51} - N_{{\rm p}_y\sigma}^{52})   &   0 &   0   &   0
    \end{pmatrix},
\end{align}
\end{subequations}
where we used the convenient shorthand notation $\{d_{xz},\,d_{yz},\,d_{x^2-y^2},\,d_{xy},\,d_{3z^2-r^2}\} \equiv \{1,\,2,\,3,\,4,\,5\}$. Similarly we find the following symmetry breaking (sb) terms
\begin{subequations}\label{eq:app_field_sb}
\begin{align}
    N_{\rm s\sigma, sb} &=
    \begin{pmatrix}
    \frac{1}{2}(N^{11}_{{\rm s}\sigma} - N^{22}_{{\rm s}\sigma})    &   \frac{1}{2}(N^{12}_{{\rm s}\sigma} + N^{21}_{{\rm s}\sigma})   &   N_{\rm s\sigma}^{13}  &   N_{\rm s\sigma}^{14}  &   N_{\rm s\sigma}^{15}\\[0.2 cm]
    \frac{1}{2}(N^{12}_{{\rm s}\sigma} + N^{21}_{{\rm s}\sigma})   &   - \frac{1}{2}(N^{11}_{{\rm s}\sigma} - N^{22}_{{\rm s}\sigma})     &   N_{\rm s\sigma}^{23}   &   N_{\rm s\sigma}^{24} &   N_{\rm s\sigma}^{25}\\[0.2 cm]
    N_{\rm s\sigma}^{31}   &   N_{\rm s\sigma}^{32}   &   0 &   0   &   N_{\rm s\sigma}^{35}\\[0.2 cm]
    N_{\rm s\sigma}^{41}   &   N_{\rm s\sigma}^{42}   &   0 &   0   &   N_{\rm s\sigma}^{45}\\[0.2 cm]
    N_{\rm s\sigma}^{51}   &   N_{\rm s\sigma}^{52}   &   N_{\rm s\sigma}^{53} &   N_{\rm s\sigma}^{54}   &   0\\[0.2 cm]
    \end{pmatrix},
\end{align}
\begin{align}
    N_{\rm d\sigma,{\rm sb}} &= 
    \begin{pmatrix}
    \frac{1}{2}(N^{11}_{{\rm d}\sigma} + N^{22}_{{\rm d}\sigma})    &   \frac{1}{2}(N^{12}_{{\rm d}\sigma} - N^{21}_{{\rm d}\sigma})   &   N^{13}_{{\rm d}\sigma}    &   N^{14}_{{\rm d}\sigma} &   N^{15}_{{\rm d}\sigma}\\[0.2 cm]
    \frac{1}{2}(N^{21}_{{\rm d}\sigma} - N^{12}_{{\rm d}\sigma})   &   \frac{1}{2}(N^{11}_{{\rm d}\sigma} + N^{22}_{{\rm d}\sigma})     &   N^{23}_{{\rm d}\sigma}   &   N^{24}_{{\rm d}\sigma} &   N^{25}_{{\rm d}\sigma}\\[0.2 cm]
    N^{31}_{{\rm d}\sigma}   &   N^{32}_{{\rm d}\sigma}   &   N^{33}_{{\rm d}\sigma}  &   N^{34}_{{\rm d}\sigma}   &   0\\[0.2 cm]
    N^{41}_{{\rm d}\sigma}    &   N^{42}_{{\rm d}\sigma}    &   N^{43}_{{\rm d}\sigma}   &   N^{44}_{{\rm d}\sigma}    &   0\\[0.2 cm]
    N^{51}_{{\rm d}\sigma}    &   N^{52}_{{\rm d}\sigma}    &   0   &   0   &   N^{55}_{{\rm d}\sigma}
    \end{pmatrix},
\end{align}
\begin{align}
    N_{{\rm p}_x\sigma, {\rm sb}} &= 
    \begin{pmatrix}
    N^{11}_{{\rm p}_x\sigma}    &   N^{12}_{{\rm p}_x\sigma}   &   \frac{1}{2}(N_{{\rm p}_x\sigma}^{13} - N_{{\rm p}_y\sigma}^{23})    &   \frac{1}{2}(N_{{\rm p}_x\sigma}^{14} - N_{{\rm p}_y\sigma}^{24})    &   \frac{1}{2}(N_{{\rm p}_x\sigma}^{15} + N_{{\rm p}_y\sigma}^{25})\\[0.2 cm]
    N^{21}_{{\rm p}_x\sigma}   &   N^{22}_{{\rm p}_x\sigma}     &   \frac{1}{2}(N_{{\rm p}_x\sigma}^{23} + N_{{\rm p}_y\sigma}^{13})   &   \frac{1}{2}(N_{{\rm p}_x\sigma}^{24} + N_{{\rm p}_y\sigma}^{14})    &   \frac{1}{2}(N_{{\rm p}_x\sigma}^{25} - N_{{\rm p}_y\sigma}^{15})\\[0.2 cm]
    \frac{1}{2}(N_{{\rm p}_x\sigma}^{31} - N_{{\rm p}_y\sigma}^{32})   &   \frac{1}{2}(N_{{\rm p}_x\sigma}^{32} + N_{{\rm p}_y\sigma}^{31})   &   N^{33}_{{\rm p}_x\sigma}   &   N^{34}_{{\rm p}_x\sigma}   &   N^{35}_{{\rm p}_x\sigma}\\[0.2 cm]
    \frac{1}{2}(N_{{\rm p}_x\sigma}^{41} - N_{{\rm p}_y\sigma}^{42})    &   \frac{1}{2}(N_{{\rm p}_x\sigma}^{42} + N_{{\rm p}_y\sigma}^{41})    &    N^{43}_{{\rm p}_x\sigma}   &   N_{{\rm p}_x\sigma}^{44}    &   N^{45}_{{\rm p}_x\sigma}\\[0.2 cm]
    \frac{1}{2}(N_{{\rm p}_x\sigma}^{51} + N_{{\rm p}_y\sigma}^{52})    &   \frac{1}{2}(N_{{\rm p}_x\sigma}^{52} - N_{{\rm p}_y\sigma}^{51})    &    N_{{\rm p}_x\sigma}^{53}    &   N^{54}_{{\rm p}_x\sigma}   &   N_{{\rm p}_x\sigma}^{55}
    \end{pmatrix},
\end{align}
\begin{align}
    N_{{\rm p}_y\sigma, {\rm sb}} &= 
    \begin{pmatrix}
    N^{11}_{{\rm p}_y\sigma}    &   N^{12}_{{\rm p}_y\sigma}   &   \frac{1}{2}(N_{{\rm p}_x\sigma}^{23} + N_{{\rm p}_y\sigma}^{13})    &   \frac{1}{2}(N_{{\rm p}_x\sigma}^{24} + N_{{\rm p}_y\sigma}^{14})    &   - \frac{1}{2}(N_{{\rm p}_x\sigma}^{25} - N_{{\rm p}_y\sigma}^{15})\\[0.2 cm]
    N^{21}_{{\rm p}_y\sigma}   &   N^{22}_{{\rm p}_y\sigma}     &   - \frac{1}{2}(N_{{\rm p}_x\sigma}^{13} - N_{{\rm p}_y\sigma}^{23})   &   - \frac{1}{2}(N_{{\rm p}_x\sigma}^{14} - N_{{\rm p}_y\sigma}^{24})    &   \frac{1}{2}(N_{{\rm p}_x\sigma}^{15} + N_{{\rm p}_y\sigma}^{25})\\[0.2 cm]
    \frac{1}{2}(N_{{\rm p}_x\sigma}^{32} + N_{{\rm p}_y\sigma}^{31})   &   - \frac{1}{2}(N_{{\rm p}_x\sigma}^{31} - N_{{\rm p}_y\sigma}^{32})   &   N^{33}_{{\rm p}_y\sigma}   &   N^{34}_{{\rm p}_y\sigma}   &   N^{35}_{{\rm p}_y\sigma}\\[0.2 cm]
    \frac{1}{2}(N_{{\rm p}_x\sigma}^{42} + N_{{\rm p}_y\sigma}^{41})    &   - \frac{1}{2}(N_{{\rm p}_x\sigma}^{41} - N_{{\rm p}_y\sigma}^{42})    &    N^{43}_{{\rm p}_y\sigma}   &   N_{{\rm p}_y\sigma}^{44}    &   N^{45}_{{\rm p}_y\sigma}\\[0.2 cm]
    - \frac{1}{2}(N_{{\rm p}_x\sigma}^{52} - N_{{\rm p}_y\sigma}^{51})    &   \frac{1}{2}(N_{{\rm p}_x\sigma}^{51} + N_{{\rm p}_y\sigma}^{52})    &    N_{{\rm p}_y\sigma}^{53}    &   N^{54}_{{\rm p}_y\sigma}   &   N_{{\rm p}_y\sigma}^{55}
    \end{pmatrix}.
\end{align}
\end{subequations}

\noindent These fields, combined with the basis functions $f_{\bf{k}}^{\gamma}$, enter in the mean-field decoupled Hamiltonian in the following way, see also Eq.~\eqref{eq:ham_mf_int}
\begin{align}
\begin{split}
    H_{\rm int} \approx H_{\rm int}^{\rm MF} = V\sum_{\bf{k}}\sum_{\mu\nu}\sum_{\sigma}c^{\,\dagger}_{\bf{k}\mu\sigma}\left[ f_{\bf{k}}^{\rm\, s}\left(N^{\mu\nu}_{{\rm s}\sigma,{\rm br}} + \alpha N^{\mu\nu}_{{\rm s}\sigma,{\rm sb}}\right) + f_{\bf{k}}^{\rm \,d}\left(N^{\mu\nu}_{{\rm d}\sigma,{\rm br}} + \alpha N^{\mu\nu}_{{\rm d}\sigma,{\rm sb}}\right)\right.\quad\quad\quad\quad
    \\ +\left. f_{\bf{k}}^{\,{\rm p}_x}\left(N^{\mu\nu}_{{\rm p}_x\sigma,{\rm br}} + \alpha N^{\mu\nu}_{{\rm p}_x\sigma,{\rm sb}}\right) + f_{\bf{k}}^{\,{\rm p}_y}\left(N^{\mu\nu}_{{\rm p}_y\sigma,{\rm br}} + \alpha N^{\mu\nu}_{{\rm p}_y\sigma,{\rm sb}}\right)\right]c^{\,\phantom{\dagger}}_{\bf{k}\nu\sigma}.
\end{split}
\end{align}
\end{widetext}

\begin{figure}[b]
    \centering
    \includegraphics[width = 0.95\columnwidth]{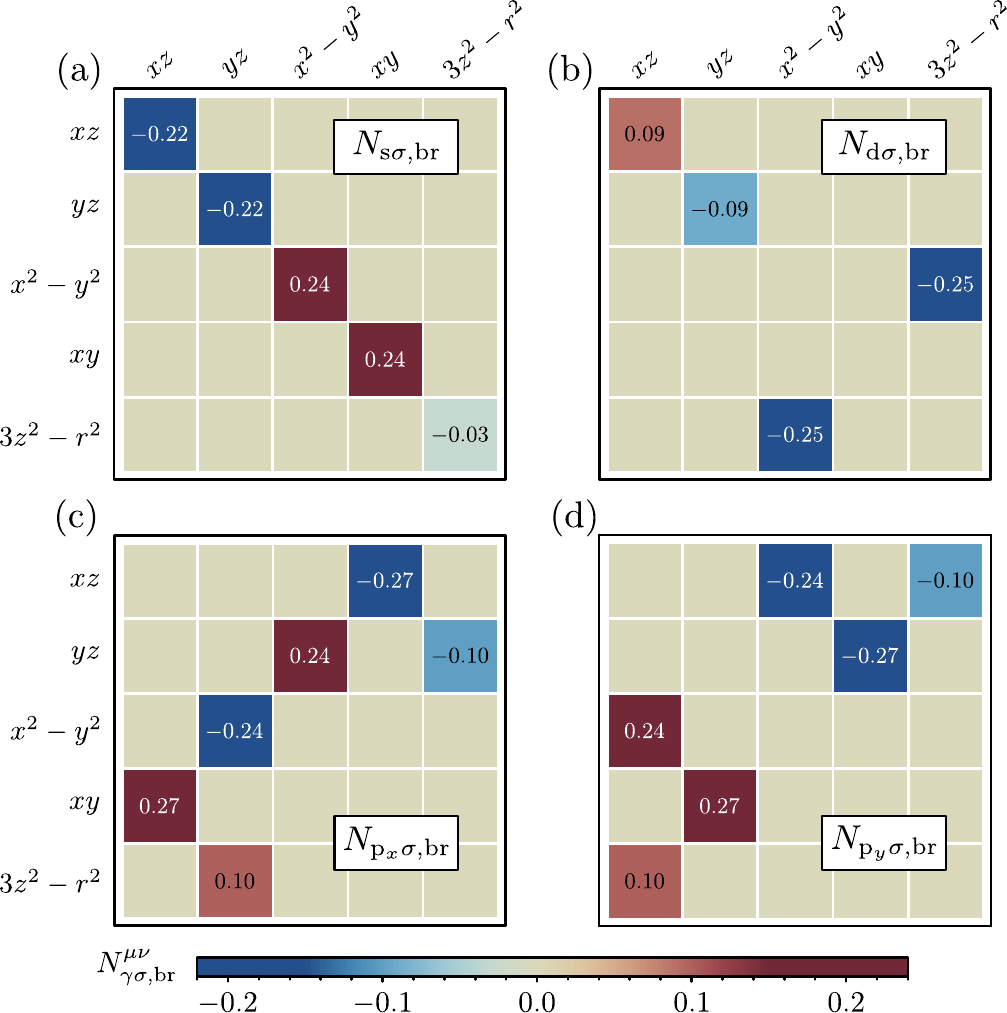}
    \caption{Band renormalizing bond-order fields, $N^{\mu\nu}_{\gamma\sigma,{\rm br}}$, calculated self-consistently with the parameters used in Fig~\ref{fig:FSs_and_band}(b) and \ref{fig:FSs_and_band}(d). A given field $N^{\mu\nu}_{\gamma\sigma,{\rm br}}$ couples to the appropriate form factor $f_{{\bf k}}^{\gamma}$, see Eq.~\eqref{eq:form_factor}. (a) and (b) are discussed in the main text, while (c) and (d) only lead to minimal effects on the low-energy band structure. All numbers appearing in the matrices have for clarity been rounded to the second decimal place.}
    \label{fig:fields_app_br}
\end{figure}
\begin{figure}[b]
    \centering
    \includegraphics[width = 0.95\columnwidth]{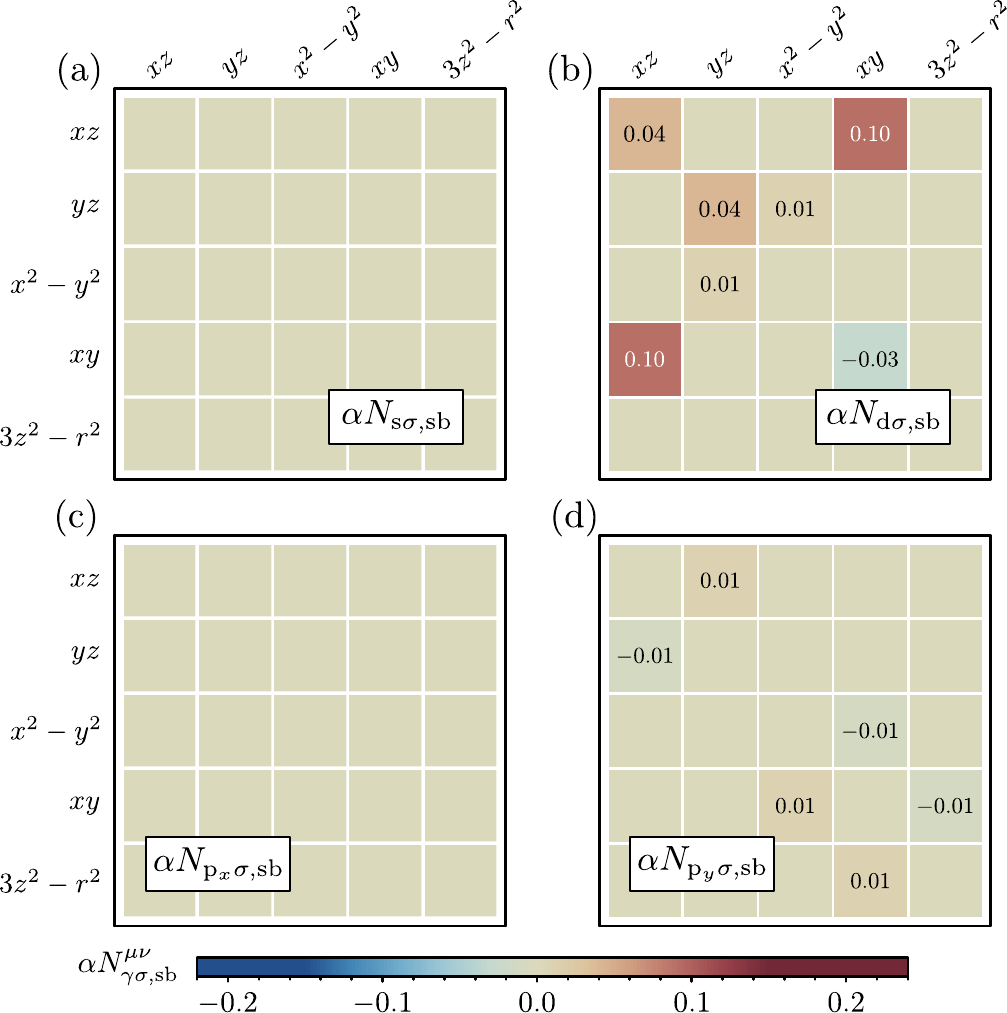}
    \caption{Symmetry breaking fields, $N_{\gamma\sigma,{\rm sb}}^{\mu\nu}$, calculated self-consistently with the parameters used in Fig~\ref{fig:FSs_and_band}(b) and \ref{fig:FSs_and_band}(d). Each field is coupled to the belonging form factor listed in Eq.~\eqref{eq:form_factor}. Only (b) and (d) display non-zero solutions, and the former will ultimately give rise to the highly anisotropic FS shown in Fig.~\ref{fig:FSs_and_band}(b). Similar to Figs.~\ref{fig:fields} and \ref{fig:fields_app_br}, also here we rounded the values of the matrix elements to the second decimal place.}
    \label{fig:fields_app_sb}
\end{figure}

The Hamiltonian becomes bilinear in creation and annihilation operators upon approximating $H_{\rm int}\approx H_{\rm int}^{\rm MF}$, and we can thus easily express the fermionic operators in the diagonal basis of the Hamiltonian $c_{{\bf k}\mu\sigma}^{\,\phantom{\dagger}}=\sum_m\gamma_{{\bf k}m\sigma}^{\,\phantom{\dagger}} \langle {\bf k}\,\mu\,\sigma| {\bf k}\,m\,\sigma\rangle \equiv \sum_m\gamma_{{\bf k}m\sigma}^{\,\phantom{\dagger}} a^{\mu\sigma}_{m}({\bf k})$, where $\gamma^{\phantom{\dagger}}_{{\bf k}m\sigma}$ are the operators related to the eigenstates of the Hamiltonian $H\gamma^{\dagger}_{{\bf k}m\sigma}|0\rangle=E_{{\bf k}m\sigma}|{\bf k}\,m\,\sigma\rangle$. In this diagonal basis the electron density $\langle n \rangle$ takes the simple form
\begin{align}
    \langle n \rangle &= \frac{1}{\mathcal{N}}\sum_{{\bf k}}\sum_{\sigma}\sum_{m}n_{\rm F}(E_{{\bf k}m\sigma}),
\end{align}
where $n_{\rm F}(E_{{\bf k}m\sigma})$ is the Fermi-Dirac distribution function. Furthermore, we find that the terms entering in the matrix elements in Eqs.~\eqref{eq:app_field_br} and \eqref{eq:app_field_sb} are expressed as
\begin{align}
\begin{split}
N^{\mu\nu}_{\gamma\sigma} &= -\frac{1}{\mathcal{N}}\sum_{{\bf k}'}\left[f^{\gamma}_{{\bf k}'}\right]^*\langle c^ {\,\dagger}_{{\bf k}'\nu\sigma} c^{\,\phantom{\dagger}}_{{\bf k}'\mu\sigma} \rangle
    \\
    &= -\frac{1}{\mathcal{N}}\sum_{{\bf k}'}\sum_m\big[f^{\gamma}_{{\bf k}'}\big]^*\,\big[a^{\nu\sigma}_m\big]^* a^{\mu\sigma}_m \, n_{\rm F}(E_{{\bf k}'m\sigma}) .
    \end{split}
\end{align}

By calculating the above self-consistently with the parameters used in Fig.~\ref{fig:FSs_and_band}(b) and \ref{fig:FSs_and_band}(c), we arrive at the band renormalizing and $S_4$ symmetry breaking fields displayed in Fig.~\ref{fig:fields_app_br} and Fig.~\ref{fig:fields_app_sb}, respectively. Only fields coupled to $f_{{\bf k}}^{\,\rm s}$ and $f_{{\bf k}}^{\,\rm d}$ are discussed in the main text, since the remaining ones lead to minimal effects on the low-energy $t_{2g}$ orbitals. This is obviously true for $N_{{\rm p}_{x}\sigma,{\rm sb}}^{\mu\nu}$ since all these are zero, while for the remaining band renormalizing terms and $N^{\mu\nu}_{{\rm p}_y\sigma,{\rm sb}}$ it is the form factors which are eliminating the effects, since $f^{\,{\rm p}_x}_{{\bf k}=(0,k_y)}=f^{\,{\rm p}_y}_{{\bf k}=(k_x,0)}=0$. This can be seen by considering the field $N^{xz\,xy}_{{\rm p}_x\sigma,{\rm br}} = -0.27$ ($N^{yz\,xy}_{{\rm p}_y\sigma,{\rm br}} = -0.27$) which introduces inter-orbital $d_{xz}-d_{xy}$ ($d_{yz}-d_{xy}$) hybridization terms, however, the form factor for this field goes to zero on the orbitally-relevant electron pocket at $Y$ ($X$), thus rendering the effect minimal. Same argument holds for $N_{{\rm p}_y\sigma,{\rm sb}}^{xz\,yz}$, which should be relevant at $\Gamma$, however $f^{\,{\rm p}_y}_{\Gamma} = 0$.

\section{Irreducible representations of bond-order fields}\label{app:ir_fields}
Although the above classification of band renormalizing and symmetry breaking fields suffices in describing the occurrence of nematic order, it fails to express the exact symmetry of the emergent nematic order. In other words, the band renormalizing and symmetry breaking fields transform as a sum of IRs, specifically
\begin{align}
    N^{\mu\nu}_{{\bf k}\sigma, \rm br} &\sim {\rm A_1} \oplus {\rm A_2}, &   N^{\mu\nu}_{{\bf k}\sigma, \rm sb} &\sim {\rm B_1} \oplus {\rm B_2} \oplus {\rm E},
\end{align}
and it is therefore not obvious whether the nematic order parameter transforms as $\rm B_1$, $\rm B_2$ or $\rm E$.

In order to shed light on this ambiguity, we will in the following perform a thorough classification of the bond-order fields, and further segregate these into IRs. In doing so, we average out the various IRs of the already established $N_{\gamma\sigma,{\rm br/sb}}^{\mu\nu}$, similar to Eq.~\eqref{eq:av_s4}, in the following way
\begin{subequations}\label{eq:av_g}
\begin{align}
    N_{\gamma, {\rm A_1}}&=\frac{1}{2}\sum_{\ell=0}^1 \mathcal{D}(C_{2}'^{\ell}) \, N_{\gamma,{\rm br}} \, \mathcal{D}^{\dagger}(C_{2}'^{\ell}),\\
    N_{\gamma,{\rm A_2}}&=N_{\gamma,{\rm br}} - N_{\gamma,{\rm A_1}},
    \\
    N_{\gamma,{\rm B_1}} &= \frac{1}{4}\sum_{\ell=0}^1\sum_{\ell'=0}^1\mathcal{D}(C_2'^{\ell'}C_2^{\ell})\,N_{\gamma,{\rm sb}}\,\mathcal{D}^{\dagger}(C_2^{\ell}C_2'^{\ell'}),
    \\
     N_{\gamma,{\rm B_2}} &= \frac{1}{2}\sum_{\ell=0}^1\mathcal{D}(C_2^{\ell})\,N_{\gamma,{\rm sb}}\,\mathcal{D}^{\dagger}(C_2^{\ell}) - N_{\gamma,{\rm B_1}},
     \\
     N_{\gamma,{\rm E}} &= N_{\gamma,{\rm sb}} - N_{\gamma,{\rm B_1}} - N_{\gamma,{\rm B_2}},
\end{align}
\end{subequations}
where $N_{\gamma,\Gamma}$ represents a matrix in combined orbital and spin space transforming as the IR $\Gamma$ of the group $\rm D_{2d}$. The $\gamma$-subscript indicates that the matrix couples to the form factor $f_{\bf k}^{\gamma}$. Similarly, $N_{\gamma,{\rm br/sb}}$ is a matrix containing the elements $N_{\gamma\sigma,{\rm br/sb}}^{\mu\nu}$. As discussed in App.~\ref{app:details_on_bond_order_fields}, $\mathcal{D}(g)$ is the matrix representation of $g\in{\rm D_{2d}}$ in combined orbital and spin space. 

By explicitly performing these averages for the 2D system under consideration, we get the following matrices $N_{\gamma\sigma,\Gamma}$ in orbital space for a given spin projection $\sigma=\{\uparrow,\downarrow\}\equiv \{1,-1\}$
\begin{widetext}
\begin{subequations}
\begin{align}
    N_{\gamma\sigma, {\rm A_1}} = \begin{pmatrix}
    \frac{1}{2}(n^{11}_{\gamma\sigma} + n^{22}_{\gamma\sigma})  & \frac{1}{2}(m^{12}_{\gamma\sigma} - m^{21}_{\gamma\sigma})  & 0 & 0  & 0  \\[0.2 cm]
    -\frac{1}{2}(m^{12}_{\gamma\sigma} - m^{21}_{\gamma\sigma})    &   \frac{1}{2}(n^{11}_{\gamma\sigma} + n^{22}_{\gamma\sigma})  &  0 & 0 & 0   \\[0.2 cm]
    0   &   0   &   n^{33}_{\gamma\sigma}    &   m^{34}_{\gamma\sigma}    &   0\\[0.2 cm]
    0   &   0   &   m^{43}_{\gamma\sigma}    &   n^{44}_{\gamma\sigma}    &   0\\[0.2 cm]
    0   &   0   &   0   &   0   &   n^{55}_{\gamma\sigma}
    \end{pmatrix},
\end{align}
\begin{align}
    N_{\gamma\sigma, {\rm A_2}} = \begin{pmatrix}
    \frac{1}{2}(m^{11}_{\gamma\sigma} + m^{22}_{\gamma\sigma})  & \frac{1}{2}(n^{12}_{\gamma\sigma} - n^{21}_{\gamma\sigma})  & 0 & 0  & 0  \\[0.2 cm]
    -\frac{1}{2}(n^{12}_{\gamma\sigma} - n^{21}_{\gamma\sigma})    &   \frac{1}{2}(m^{11}_{\gamma\sigma} + m^{22}_{\gamma\sigma})  &  0 & 0 & 0   \\[0.2 cm]
    0   &   0   &   m^{33}_{\gamma\sigma}    &   n^{34}_{\gamma\sigma}    &   0\\[0.2 cm]
    0   &   0   &   n^{43}_{\gamma\sigma}    &   m^{44}_{\gamma\sigma}    &   0\\[0.2 cm]
    0   &   0   &   0   &   0   &   m^{55}_{\gamma\sigma}
    \end{pmatrix},
\end{align}
\begin{align}
    N_{\gamma\sigma, {\rm B_1}} = \begin{pmatrix}
    \frac{1}{2}(n^{11}_{\gamma\sigma} - n^{22}_{\gamma\sigma})  & \frac{1}{2}(m^{12}_{\gamma\sigma} + m^{21}_{\gamma\sigma})  & 0 & 0  & 0    \\[0.2 cm]
    \frac{1}{2}(m^{12}_{\gamma\sigma} + m^{21}_{\gamma\sigma})    &   -\frac{1}{2}(n^{11}_{\gamma\sigma} - n^{22}_{\gamma\sigma})  &  0 & 0 & 0   \\[0.2 cm]
    0   &   0   &   0    &   0    &   n^{35}_{\gamma\sigma}\\[0.2 cm]
    0   &   0   &   0    &   0    &   m^{45}_{\gamma\sigma}\\[0.2 cm]
    0   &   0   &   n^{53}_{\gamma\sigma}   &   m^{54}_{\gamma\sigma}   &   0
    \end{pmatrix},
\end{align}
\begin{align}
    N_{\gamma \sigma, {\rm B_2}} = \begin{pmatrix}
    \frac{1}{2}(m^{11}_{\gamma\sigma} - m^{22}_{\gamma\sigma})  & \frac{1}{2}(n^{12}_{\gamma\sigma} + n^{21}_{\gamma\sigma})  & 0 & 0  & 0  \\[0.2 cm]
    \frac{1}{2}(n^{12}_{\gamma\sigma} + n^{21}_{\gamma\sigma})    &   -\frac{1}{2}(m^{11}_{\gamma\sigma} - m^{22}_{\gamma\sigma})  &  0 & 0 & 0   \\[0.2 cm]
    0   &   0   &   0    &   0    &   m^{35}_{\gamma\sigma}\\[0.2 cm]
    0   &   0   &   0    &   0    &   n^{45}_{\gamma\sigma}\\[0.2 cm]
    0   &   0   &   m^{53}_{\gamma\sigma}   &   n^{54}_{\gamma\sigma}   &   0
    \end{pmatrix},
\end{align}
\begin{align}
    N_{\gamma \sigma, {\rm E}_x} = \begin{pmatrix}
    0  & 0  & \frac{1}{2}(m_{\gamma\sigma}^{13} + m_{\gamma\sigma}^{23}) & \frac{1}{2}(n_{\gamma\sigma}^{14} + n_{\gamma\sigma}^{24})  & \frac{1}{2}(m_{\gamma\sigma}^{15} - m_{\gamma\sigma}^{25})  \\[0.2 cm]
    0    &   0  &  \frac{1}{2}(- n_{\gamma\sigma}^{13} + n_{\gamma\sigma}^{23}) & \frac{1}{2}(- m_{\gamma\sigma}^{14} + m_{\gamma\sigma}^{24}) & \frac{1}{2}( n_{\gamma\sigma}^{15} + n_{\gamma\sigma}^{25})   \\
    \frac{1}{2}( m_{\gamma\sigma}^{31} + m_{\gamma\sigma}^{32})   &   \frac{1}{2}(- n_{\gamma\sigma}^{31} + n_{\gamma\sigma}^{32})   &   0    &   0    &   0\\[0.2 cm]
    \frac{1}{2}(n_{\gamma\sigma}^{41} + n_{\gamma\sigma}^{42})   &   \frac{1}{2}(- m_{\gamma\sigma}^{41} + m_{\gamma\sigma}^{42})   &   0    &   0    &   0\\[0.2 cm]
    \frac{1}{2}(m_{\gamma\sigma}^{51} - m_{\gamma\sigma}^{52})   &   \frac{1}{2}( n_{\gamma\sigma}^{51} + n_{\gamma\sigma}^{52})   &   0   &   0   &   0
    \end{pmatrix},
\end{align}
\begin{align}
    N_{\gamma \sigma, {\rm E}_y} = \begin{pmatrix}
    0  & 0  & \frac{1}{2}(n_{\gamma\sigma}^{13} - n_{\gamma\sigma}^{23}) & \frac{1}{2}(m_{\gamma\sigma}^{14} - m_{\gamma\sigma}^{24})  & \frac{1}{2}(n_{\gamma\sigma}^{15} + n_{\gamma\sigma}^{25})  \\[0.2 cm]
    0    &   0  &  \frac{1}{2}(m_{\gamma\sigma}^{13} + m_{\gamma\sigma}^{23}) & \frac{1}{2}(n_{\gamma\sigma}^{14} + n_{\gamma\sigma}^{24}) & \frac{1}{2}( - m_{\gamma\sigma}^{15} + m_{\gamma\sigma}^{25})   \\[0.2 cm]
    \frac{1}{2}( n_{\gamma\sigma}^{31} - n_{\gamma\sigma}^{32})   &   \frac{1}{2}(m_{\gamma\sigma}^{31} + m_{\gamma\sigma}^{32})   &   0    &   0    &   0\\[0.2 cm]
    \frac{1}{2}(m_{\gamma\sigma}^{41} - m_{\gamma\sigma}^{42})   &   \frac{1}{2}( n_{\gamma\sigma}^{41} + n_{\gamma\sigma}^{42})   &   0    &   0    &   0\\[0.2 cm]
    \frac{1}{2}(n_{\gamma\sigma}^{51} + n_{\gamma\sigma}^{52})   &   \frac{1}{2}( - m_{\gamma\sigma}^{51} + m_{\gamma\sigma}^{52})   &   0   &   0   &   0
    \end{pmatrix},
\end{align}
\begin{align}
    N_{\gamma \sigma, {\rm E}_{L_x}} = \begin{pmatrix}
    0  & 0  & \frac{1}{2}(m_{\gamma\sigma}^{13} - m_{\gamma\sigma}^{23}) & \frac{1}{2}(n_{\gamma\sigma}^{14} - n_{\gamma\sigma}^{24})  & \frac{1}{2}(m_{\gamma\sigma}^{15} + m_{\gamma\sigma}^{25})  \\[0.2 cm]
    0    &   0  &  \frac{1}{2}(n_{\gamma\sigma}^{13} + n_{\gamma\sigma}^{23}) & \frac{1}{2}(m_{\gamma\sigma}^{14} + m_{\gamma\sigma}^{24}) & \frac{1}{2}( - n_{\gamma\sigma}^{15} + n_{\gamma\sigma}^{25})   \\[0.2 cm]
    \frac{1}{2}( m_{\gamma\sigma}^{31} - m_{\gamma\sigma}^{32})   &   \frac{1}{2}( n_{\gamma\sigma}^{31} + n_{\gamma\sigma}^{32})   &   0    &   0    &   0\\[0.2 cm]
    \frac{1}{2}(n_{\gamma\sigma}^{41} - n_{\gamma\sigma}^{42})   &   \frac{1}{2}( m_{\gamma\sigma}^{41} + m_{\gamma\sigma}^{42})   &   0    &   0    &   0\\[0.2 cm]
    \frac{1}{2}(m_{\gamma\sigma}^{51} + m_{\gamma\sigma}^{52})   &   \frac{1}{2}( - n_{\gamma\sigma}^{51} + n_{\gamma\sigma}^{52})   &   0   &   0   &   0
    \end{pmatrix},
\end{align}
\begin{align}
    N_{\gamma \sigma, {\rm E}_{L_y}} = \begin{pmatrix}
    0  & 0  & \frac{1}{2}(n_{\gamma\sigma}^{13} + n_{\gamma\sigma}^{23}) & \frac{1}{2}(m_{\gamma\sigma}^{14} + m_{\gamma\sigma}^{24})  & \frac{1}{2}(n_{\gamma\sigma}^{15} - n_{\gamma\sigma}^{25})  \\[0.2 cm]
    0    &   0  &  \frac{1}{2}(- m_{\gamma\sigma}^{13} + m_{\gamma\sigma}^{23}) & \frac{1}{2}(- n_{\gamma\sigma}^{14} + n_{\gamma\sigma}^{24}) & \frac{1}{2}( m_{\gamma\sigma}^{15} + m_{\gamma\sigma}^{25})   \\
    \frac{1}{2}( n_{\gamma\sigma}^{31} + n_{\gamma\sigma}^{32})   &   \frac{1}{2}(- m_{\gamma\sigma}^{31} + m_{\gamma\sigma}^{32})   &   0    &   0    &   0\\[0.2 cm]
    \frac{1}{2}(m_{\gamma\sigma}^{41} + m_{\gamma\sigma}^{42})   &   \frac{1}{2}( - n_{\gamma\sigma}^{41} + n_{\gamma\sigma}^{42})   &   0    &   0    &   0\\[0.2 cm]
    \frac{1}{2}(n_{\gamma\sigma}^{51} - n_{\gamma\sigma}^{52})   &   \frac{1}{2}( m_{\gamma\sigma}^{51} + m_{\gamma\sigma}^{52})   &   0   &   0   &   0
    \end{pmatrix},
\end{align}
\end{subequations}
\end{widetext}
where we introduced the following quantities for brevity
\begin{align}
    n_{\gamma\sigma}^{\mu\nu} &= (N_{\gamma\sigma}^{\mu\nu} + N_{\gamma\bar{\sigma}}^{\mu\nu})/2, &
    m_{\gamma\sigma}^{\mu\nu} &= \sigma(N_{\gamma\sigma}^{\mu\nu} - N_{\gamma\bar{\sigma}}^{\mu\nu})/2,
\end{align}
with $\sigma = -\bar{\sigma}$. Note furthermore that the matrices labeled by the IR ${\rm E}_{x,y}$ transform as the basis functions $(x,y)$, while $N_{\gamma\sigma,{\rm E}_{L_{x}}}$ and $N_{\gamma\sigma,{\rm E}_{L_{y}}}$ instead transform as the angular momentum pseudovector $(L_{x}, L_y)$. In general we allow for magnetic terms, i.e. $m_{\gamma\sigma}^{\mu\nu}\neq 0$, however, we find these fields to be orders of magnitude smaller than the density terms $n^{\mu\nu}_{\gamma\sigma}$. Nonetheless, for completeness our calculations and classification include all terms, so not to miss any subtle details.

We can then, after having performed the various averages in Eq.~\eqref{eq:av_g}, express $N_{{\bf k}\sigma}^{\mu\nu}$ in terms of the IRs in the following way
\begin{align}
\begin{split}
    N_{{\bf k}\sigma}^{\mu\nu}
    &= \sum_{\gamma}f_{\bf k}^{\gamma}\big(N_{\gamma\sigma,{\rm br}}^{\mu\nu} + N_{\gamma\sigma,{\rm sb}}^{\mu\nu}\big) = \sum_{\gamma,\Gamma}f_{\bf k}^{\gamma} N_{\gamma\sigma,\Gamma}^{\mu\nu}.
    \end{split}
\end{align}
This informs us that a single term $f_{\bf k}^{\gamma} N_{\gamma\sigma,\Gamma}^{\mu\nu}$ must transform as the product of the two IRs $\gamma$ and $\Gamma$, e.g. for $\gamma = {\rm d}\sim {\rm B_1}$ and $\Gamma = {\rm A_1}$, the term altogether transforms as $\rm B_1 \otimes A_1 = \rm B_1$. With this in mind, we can collect terms that transform equivalently, and get
\begin{align}\label{eq:A1_terms}
    &N_{{\bf k}\sigma,\rm A_1}^{\mu\nu} = f_{\bf k}^{\,\rm s}N_{{\rm s}\sigma,{\rm A_1}}^{\mu\nu} + f_{\bf k}^{\,\rm d}N_{{\rm d}\sigma,{\rm B_1}}^{\mu\nu} \\&+ f_{\bf k}^{{\rm p}_x}\frac{N_{{\rm p}_x\sigma,{\rm E}_x} + N_{{\rm p}_x\sigma,{\rm E}_{L_x}}}{2} + f_{\bf k}^{{\rm p}_y}\frac{N_{{\rm p}_x\sigma,{\rm E}_y} - N_{{\rm p}_x\sigma,{\rm E}_{L_y}}}{2}\nonumber\\&
    +f_{\bf k}^{{\rm p}_x}\frac{N_{{\rm p}_y\sigma,{\rm E}_x} - N_{{\rm p}_y\sigma,{\rm E}_{L_x}}}{2} + f_{\bf k}^{{\rm p}_y}\frac{N_{{\rm p}_y\sigma,{\rm E}_y} + N_{{\rm p}_y\sigma,{\rm E}_{L_y}}}{2},\nonumber
\end{align}
\begin{align}
    &N^{\mu\nu}_{{\bf k}\sigma,\rm A_2} = f_{\bf k}^{\,\rm s}N_{{\rm s}\sigma,{\rm A_2}}^{\mu\nu} + f_{\bf k}^{\,\rm d}N_{{\rm d}\sigma,{\rm B_2}}^{\mu\nu} \\&+ f_{\bf k}^{{\rm p}_x}\frac{N_{{\rm p}_x\sigma,{\rm E}_y} + N_{{\rm p}_x\sigma,{\rm E}_{L_y}}}{2} - f_{\bf k}^{{\rm p}_y}\frac{N_{{\rm p}_x\sigma,{\rm E}_x} - N_{{\rm p}_x\sigma,{\rm E}_{L_x}}}{2}\nonumber\\
    &-f_{\bf k}^{{\rm p}_x}\frac{N_{{\rm p}_y\sigma,{\rm E}_y} - N_{{\rm p}_y\sigma,{\rm E}_{L_y}}}{2} + f_{\bf k}^{{\rm p}_y}\frac{N_{{\rm p}_y\sigma,{\rm E}_x} + N_{{\rm p}_y\sigma,{\rm E}_{L_x}}}{2},\nonumber
\end{align}
\begin{align}\label{eq:B1_terms}
    &N^{\mu\nu}_{{\bf k}\sigma,\rm B_1} = f_{\bf k}^{\,\rm s}N_{{\rm s}\sigma,{\rm B_1}}^{\mu\nu} + f_{\bf k}^{\,\rm d}N_{{\rm d}\sigma,{\rm A_1}}^{\mu\nu}\\&+ f_{\bf k}^{{\rm p}_x}\frac{N_{{\rm p}_x,\sigma{\rm E}_x} + N_{{\rm p}_x\sigma,{\rm E}_{L_x}}}{2} - f_{\bf k}^{{\rm p}_y}\frac{N_{{\rm p}_x\sigma,{\rm E}_y} - N_{{\rm p}_x\sigma,{\rm E}_{L_y}}}{2}\nonumber\\
    &- f_{\bf k}^{{\rm p}_x}\frac{N_{{\rm p}_y\sigma,{\rm E}_x} - N_{{\rm p}_y\sigma,{\rm E}_{L_x}}}{2} + f_{\bf k}^{{\rm p}_y}\frac{N_{{\rm p}_y,\sigma{\rm E}_y} + N_{{\rm p}_y,\sigma{\rm E}_{L_y}}}{2},\nonumber
\end{align}
\begin{align}
    &N^{\mu\nu}_{{\bf k}\sigma,\rm B_2} = f_{\bf k}^{\,\rm s}N_{{\rm s}\sigma,{\rm B_2}}^{\mu\nu} + f_{\bf k}^{\,\rm d}N_{{\rm d}\sigma,{\rm A_2}}^{\mu\nu}\\&+ f_{\bf k}^{{\rm p}_x}\frac{N_{{\rm p}_x\sigma,{\rm E}_y} + N_{{\rm p}_x\sigma,{\rm E}_{L_y}}}{2} + f_{\bf k}^{{\rm p}_y}\frac{N_{{\rm p}_x\sigma,{\rm E}_x} - N_{{\rm p}_x\sigma,{\rm E}_{L_x}}}{2}\nonumber\\
    &+f_{\bf k}^{{\rm p}_x}\frac{N_{{\rm p}_y\sigma,{\rm E}_y} - N_{{\rm p}_y\sigma,{\rm E}_{L_y}}}{2} + f_{\bf k}^{{\rm p}_y}\frac{N_{{\rm p}_y\sigma,{\rm E}_x} + N_{{\rm p}_y\sigma,{\rm E}_{L_x}}}{2},\nonumber
\end{align}
for the IRs $\rm A_1$, $\rm A_2$, $\rm B_1$ and $\rm B_2$, respectively, and finally for the 2D IR we arrive at
\begin{align}\label{eq:IR_E}
    &N_{{\bf k}\sigma,\rm E}^{\mu\nu} = f_{\bf k}^{\,\rm s}(N_{{\rm s}\sigma,{\rm E}_x}^{\mu\nu} + N_{{\rm s}\sigma,{\rm E}_y}^{\mu\nu}) + f_{\bf k}^{\,\rm s}(N_{{\rm s}\sigma,{\rm E}_{L_x}}^{\mu\nu} + N_{{\rm s}\sigma,{\rm E}_{L_y}}^{\mu\nu})\nonumber\\
    &+ f_{\bf k}^{\,\rm d}(N_{{\rm d}\sigma,{\rm E}_x}^{\mu\nu} + N_{{\rm d}\sigma,{\rm E}_y}^{\mu\nu}) +f_{\bf k}^{\,\rm d}(N_{{\rm d}\sigma,{\rm E}_{L_x}}^{\mu\nu} + N_{{\rm d}\sigma,{\rm E}_{L_y}}^{\mu\nu})\nonumber
    \\
    &+f_{\bf k}^{\,{\rm p}_x}(N_{{\rm p}_x\sigma,{\rm A}_1}^{\mu\nu} + N_{{\rm p}_x\sigma,{\rm B}_1}^{\mu\nu} + N_{{\rm p}_x\sigma,{\rm A}_2}^{\mu\nu} + N_{{\rm p}_x\sigma,{\rm B}_2}^{\mu\nu})\nonumber\\ &+ 
    f_{\bf k}^{\,{\rm p}_y}(N_{{\rm p}_y\sigma,{\rm A}_1}^{\mu\nu} + N_{{\rm p}_y\sigma,{\rm B}_1}^{\mu\nu} + N_{{\rm p}_y\sigma,{\rm A}_2}^{\mu\nu} + N_{{\rm p}_y\sigma,{\rm B}_2}^{\mu\nu}).
\end{align}
From the above, we are able to pinpoint exactly what fields give rise to the nematic order. For example, if a non-zero $\rm B_1$ term appears in our self-consistent calculations, then we know that the nematic order transforms as $\rm B_1$, and that the resulting crystalline point group must be $\rm D_2 \vartriangleleft \rm D_{2d}$. 

\begin{figure}
    \centering
    \includegraphics[width = 0.95 \columnwidth]{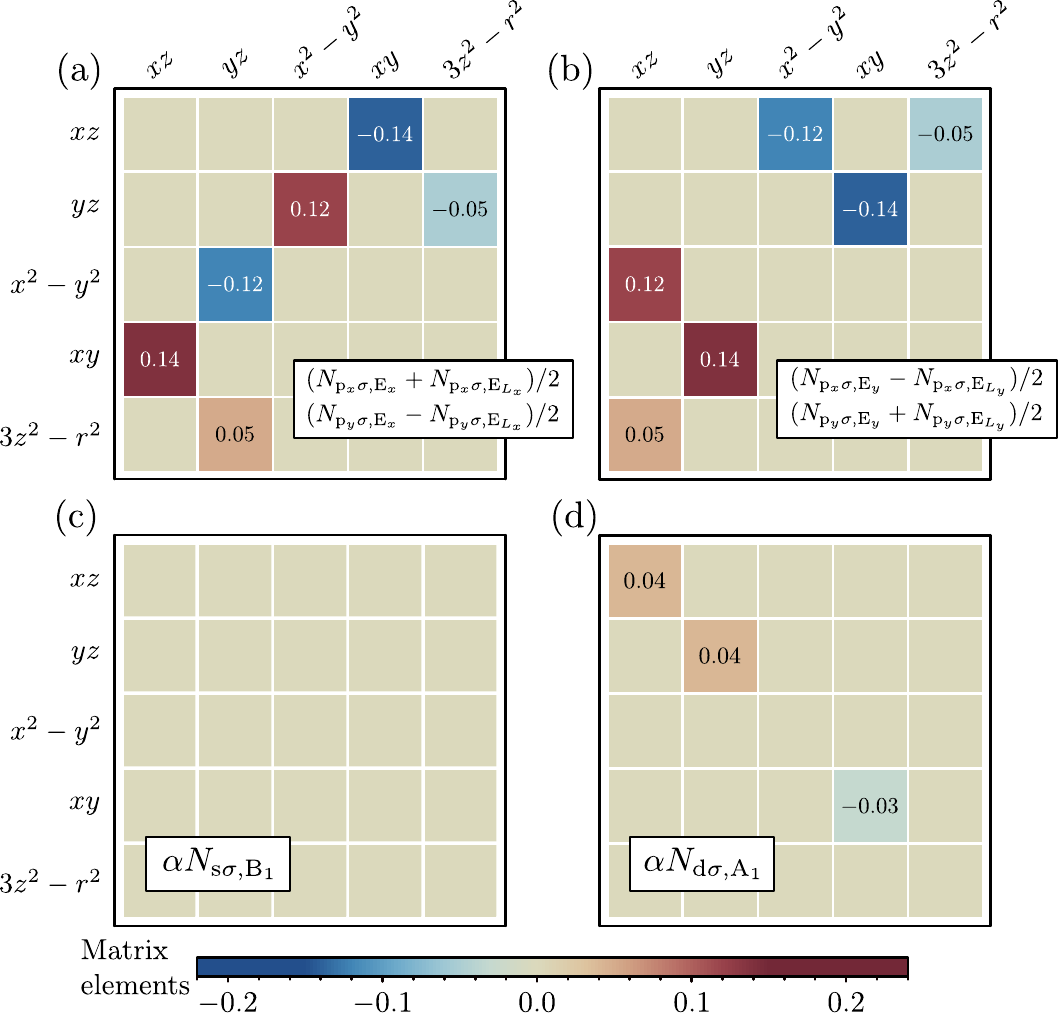}
    \caption{(a) and (b) Matrix structure of fields coupled to the p-wave form factors $f_{\bf k}^{{\rm p}_x}$ and $f_{\bf k}^{{\rm p}_y}$, respectively. These fields enter both in $N_{{\bf k} \sigma,{\rm A_1}}$ and $N_{{\bf k} \sigma,{\rm B_1}}$, but only lead to a non-zero contribution in the former, see Eqs.~\eqref{eq:A1_terms} and \eqref{eq:B1_terms}. (c) and (d) Remaining contributions entering in $N_{{\bf k}\sigma,{\rm B_1}}$. The non-zero elements in (d), can be utilized as the nematic order parameter, since they break $S_4$ symmetry. See Eq.~\eqref{eq:B1_op} for the resulting B$_1$ neamtic order parameter. All values in the figure have been rounded to the second decimal place.}
    \label{fig:fields_app_B1}
\end{figure}

From our self-consistent calculations, we find that all A$_2$ and B$_2$ terms are identically zero, i.e. $N^{\mu\nu}_{{\bf k}\sigma,{\rm A_2}} = N^{\mu\nu}_{{\bf k}\sigma,{\rm B_2}}=0$, which immediately implies that $N^{\mu\nu}_{{\bf k}\sigma,{\rm br}} \equiv N^{\mu\nu}_{{\bf k}\sigma,{\rm A_1}}$, thus the band renormalizing terms all transform as the IR A$_1$. See Fig.~\ref{fig:fields_app_br} for the values of $N_{{\bf k}\sigma,{\rm A_1}}^{\mu\nu}$, where we straightforwardly conclude that
\begin{align}
   N_{{\rm s}\sigma,{\rm br}}^{\mu\nu} &\equiv N_{{\rm s}\sigma,{\rm A_1}}^{\mu\nu},   &   N_{{\rm d}\sigma,{\rm br}}^{\mu\nu}&\equiv N_{{\rm d}\sigma,{\rm B_1}}^{\mu\nu}.
\end{align}
Similarly we conclude   
\begin{align}
\begin{split}
    N_{{\rm p}_x\sigma,{\rm br}}^{\mu\nu} &\equiv \frac{N_{{\rm p}_x\sigma,{\rm E}_x}^{\mu\nu} + N_{{\rm p}_x\sigma,{\rm E}_{L_x}}^{\mu\nu} + N_{{\rm p}_y\sigma,{\rm E}_x}^{\mu\nu} - N_{{\rm p}_x\sigma,{\rm E}_{L_x}}^{\mu\nu}}{2}, 
    \\
    N_{{\rm p}_y\sigma,{\rm br}}^{\mu\nu} &\equiv \frac{N_{{\rm p}_x\sigma,{\rm E}_y}^{\mu\nu} - N_{{\rm p}_x\sigma,{\rm E}_{L_y}}^{\mu\nu} + N_{{\rm p}_y\sigma,{\rm E}_y}^{\mu\nu} + N_{{\rm p}_x\sigma,{\rm E}_{L_y}}^{\mu\nu}}{2},
    \end{split}
\end{align}
which can be inferred from Fig.~\ref{fig:fields_app_B1}(a) and (b), where we show the matrices with the form factors $f_{\rm k}^{\,{\rm p}_{x}}$ and $f_{\rm k}^{\,{\rm p}_{y}}$ in Eq.~\eqref{eq:A1_terms}.

We can furthermore extract from Fig.~\ref{fig:fields_app_B1}(a) and (b) that $N_{{\bf k}\sigma,{\rm B_1}}^{\mu\nu}$ only involve terms coupling to $f_{\bf k}^{\rm s}$ and $f_{\bf k}^{\rm d}$, since
\begin{align}
\begin{split}
    \frac{N_{{\rm p}_x\sigma,{\rm E}_x}^{\mu\nu} + N_{{\rm p}_x\sigma,{\rm E}_{L_x}}^{\mu\nu}}{2} - \frac{N_{{\rm p}_y\sigma,{\rm E}_x}^{\mu\nu} - N_{{\rm p}_x\sigma,{\rm E}_{L_x}}^{\mu\nu}}{2} = 0
    \\
    -\frac{N_{{\rm p}_x\sigma,{\rm E}_y}^{\mu\nu} - N_{{\rm p}_x\sigma,{\rm E}_{L_y}}^{\mu\nu}}{2} + \frac{N_{{\rm p}_y\sigma,{\rm E}_y}^{\mu\nu} + N_{{\rm p}_x\sigma,{\rm E}_{L_y}}^{\mu\nu}}{2} = 0,
    \end{split}
\end{align}
i.e. all terms coupling to $f_{\bf k}^{\,{\rm p}_{x,y}}$ cancel in this IR channel. From our self-consistent calculations, we also find that $N_{\rm s\sigma,{\rm B_1}}^{\mu\nu}=0$, leaving us with only a single remaining term in $N_{{\bf k}\sigma, {\rm B_1}}^{\mu\nu}$, namely $N^{\mu\nu}_{{\rm d\sigma,{\rm A_1}}}$. See Fig.~\ref{fig:fields_app_B1}(c) and (d) for an illustation of the matrices $N_{{\rm s}\sigma,{\rm B_1}}$ and $N_{{\rm d}\sigma,{\rm A_1}}$. The non-zero matrix elements of the latter can serve as order parameters for our nematic phase, since these break $S_4$ symmetry and reduce the crystalline point group symmetry to the group $\rm D_{2}$. Specifically, we define the $\rm B_1$ nematic order parameter as
\begin{align}\label{eq:B1_op}
\begin{split}
    N_{\rm B_1} &= (N_{\rm d\sigma,A_1}^{11} + N_{\rm d\sigma,A_1}^{22}) / 2
    \\
    &= (N_{\rm d\sigma}^{11} + N_{\rm d\bar{\sigma}}^{11} + N_{\rm d\sigma}^{22} + N_{\rm d\bar{\sigma}}^{22}) / 4
    \\
    &= (N_{\rm d\sigma,{\rm sb}}^{11} + N_{\rm d\bar{\sigma},{\rm sb}}^{11} + N_{\rm d\sigma,{\rm sb}}^{22} + N_{\rm d\bar{\sigma},{\rm sb}}^{22}) / 4.
    \end{split}
\end{align}
Alternatively, one could also define the order parameter as $N^{44}_{{\rm d}\sigma,{\rm A_1}}$, however, we choose to focus on $N_{\rm B_1}$ since it acquires a slightly higher value.

\begin{figure}
    \centering
    \includegraphics[width = 0.95 \columnwidth]{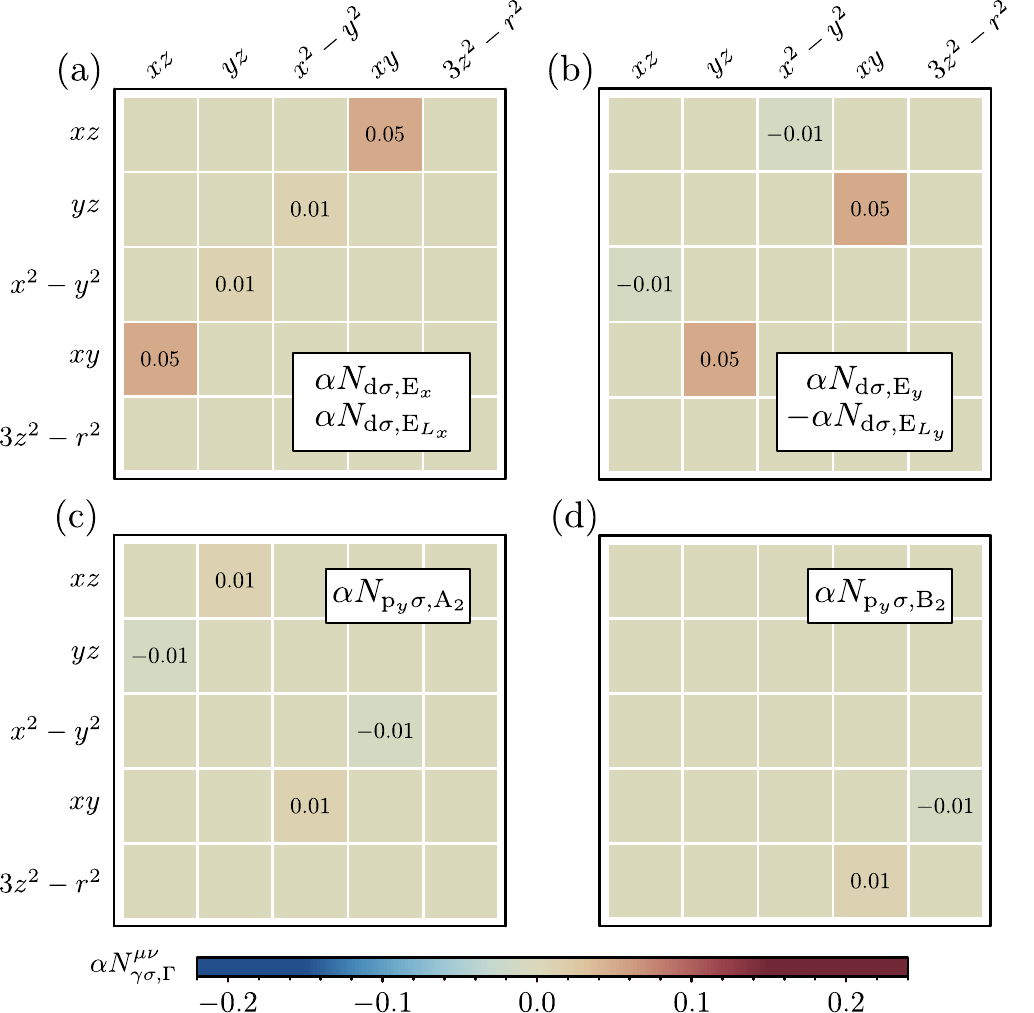}
    \caption{Non-zero matrices entering in the 2D IR fields in Eq.~\eqref{eq:IR_E}. The matrices in (a) and (b) [(c) and (d)] couple to the form factor $f^{\,\rm d}_{\bf k}$ ($f^{\,{\rm p}_y}_{\bf k}$). As for the B$_1$ field, also here the non-zero elements allow us to define an order parameter, see Eq.~\eqref{eq:op_E}, which can be used as an indicator for when the symmetries of the system are described by the monoclinic point group $\rm C_2$. All values in the figure have been rounded to the second decimal place.}
    \label{fig:fields_app_E}
\end{figure}

For the terms transforming as the 2D IR $\rm E$, we find the non-zero matrices displayed in Fig.~\ref{fig:fields_app_E}, which ultimately result in a simplified expression for $N_{{\bf k}\sigma,{\rm E}}$, namely
\begin{align}
    N_{{\bf k}\sigma,{\rm E}}^{\mu\nu}&=
    \\
    f_{\bf k}^{\,\rm d}&(N_{{\rm d}\sigma,{\rm E}_x}^{\mu\nu} + N_{{\rm d}\sigma,{\rm E}_{L_x}}^{\mu\nu})
    + f_{\bf k}^{\,{\rm p}_y}(N_{{\rm p}_y\sigma,{\rm A}_2}^{\mu\nu} + N_{{\rm p}_y\sigma,{\rm B}_2}^{\mu\nu}).\nonumber
\end{align}
Similar to the B$_1$ terms, also here we can adopt the non-zero elements of $N^{\mu\nu}_{{\bf k}\sigma,{\rm E}}$ as an order parameter of the system. Specifically we define the following two component order parameter
\begin{align}\label{eq:op_E}
\begin{split}
    N_{\rm E} &= \left(N^{14}_{{\rm d}\sigma,{\rm E}_x} + N^{14}_{{\rm d}\sigma,{\rm E}_{L_x}}, N^{24}_{{\rm d}\sigma,{\rm E}_y} + N^{24}_{{\rm d}\sigma,{\rm E}_{L_y}}\right)
    \\
    &= \left([N_{{\rm d}\sigma}^{14} + N_{{\rm d}\bar{\sigma}}^{14}] / 2, [N_{{\rm d}\sigma}^{24} + N_{{\rm d}\bar{\sigma}}^{24}] / 2\right)
    \\
    &= \left([N_{{\rm d}\sigma,{\rm sb}}^{14} + N_{{\rm d}\bar{\sigma},{\rm sb}}^{14}] / 2, [N_{{\rm d}\sigma,{\rm sb}}^{24} + N_{{\rm d}\bar{\sigma},{\rm sb}}^{24}] / 2\right)
    \\
    &\equiv (N_{{\rm E}_x}, N_{{\rm E}_y}).
    \end{split}
\end{align}
The order parameters $N_{{\rm E}_{x,y}}$ will enter on equal footing in a Landau free energy expansion, since they are interrelated by symmetry, and the system can thus display either $N_{{\rm E}_x}\neq 0$ or $N_{{\rm E}_y}\neq 0$. Through our self-consistent calculations we find in fact both solutions, depending on the initialization of our computations, i.e. where in the energy landscape the calculations start. Throughout the paper we have focused on solutions with $N_{{\rm E}_x}\neq 0$.

Conclusively, we see that the presence of two non-zero order parameters, belonging to distinct IRs, hints at the following two phase transition scenarios: $i)$ By lowering of the temperature, $N_{\rm E}$ becomes non-zero which imposes the symmetry point group transition $\rm D_{2d}\mapsto {\rm C}_2$. For this specific point group $N_{\rm B_1}$ transform trivially, and is thus allowed to enter simultaneously with $N_{\rm E}$, i.e. $N_{\rm E}$ and $N_{\rm B}$ emerges at the same transition temperature. The other scenario is $ii)$ The system undergoes two consecutive phase transitions, with $N_{\rm B_1}$ entering prior to $N_{\rm E}$ when lowering the temperature, leading to two transition temperatures and the following point group reduction scheme ${\rm D}_{\rm 2d}\mapsto{\rm D}_{\rm 2}\mapsto{\rm C}_{\rm 2}$. For the spicific system under consideration, we find scenario $ii)$ to be true, see Fig.~\ref{fig:temp_sweep} where we display the order parameters $N_{\rm B}$ and $N_{\rm E}$ at various temperatures. Nonetheless, we stress that case $i)$ potentially also could arise in experiments.

\section{Susceptibility and spin-fluctuation pairing from a partially incoherent electronic structure}\label{app:susc}
\renewcommand{\k}{{\bf k}}
\newcommand{\q}{{\bf q}}
Adopting the properties of a correlated electron gas which is characterized by 
a reduced quasiparticle weight $Z({\bf k}_\mathrm{F})$ at the Fermi level, it seems a good approximation
at low energies to parametrize the Green function as
\begin{align}\label{eq_GF}
 G_{\mu\nu} (\k,\omega_n)%&=\sqrt{Z_\ell  Z_{\ell'}} \;\sum_\mu \frac{a_\mu^\ell(\k) a_\mu^{\ell'*}(\k)}{i \omega_n - \tilde E_\mu(\k)}\notag\\
&=\sqrt{Z_{\mu}  Z_{\nu}} \;\sum_m a_m^\mu(\k) a_m^{\nu*}(\k) G^m(\k,\omega_n),
\end{align}
where $Z_\mu$ are quasiparticle weights in orbital $\mu$, $E_{\k m}$ are the eigenenergies of band $m$ of the mean field Hamiltonian in Eq. (\ref{eq:ham_mf_int}), $a_m^\mu(\k)$ the corresponding orbital component of the eigenstate and $ G^m(\k,\omega_n)=[i \omega_n -  E_{\k m}]^{-1}$ the Green function of band $m$.
Adopting the usual local interactions from the Hubbard-Hund Hamiltonian  
\begin{eqnarray}
	H &=& {U}\sum_{i,\mu}n_{i\mu\uparrow}n_{i\mu\downarrow}+{U}'\sum_{i,\nu<\mu}n_{i\mu}n_{i\nu}
	\nonumber\\
	& + & {J}\sum_{i,\nu<\mu}\sum_{\sigma,\sigma'}c_{i\mu\sigma}^{\dagger}c_{i\nu\sigma'}^{\dagger}c_{i\mu\sigma'}c_{i\nu\sigma}\nonumber\\
	& + & {J}'\sum_{i,\nu\neq\mu}c_{i\mu\uparrow}^{\dagger}c_{i\mu\downarrow}^{\dagger}c_{i\nu\downarrow}c_{i\nu\uparrow} , \label{H_int}
\end{eqnarray}
where the parameters ${U}$, ${U}'$, ${J}$, ${J}'$ are given in the notation of Kuroki \textit{et al.} \cite{Kuroki2008}, we stay in the regime where $U'=U-2J$, $J=J'$ and use the overall interaction magnitude $U$  as free parameter to tune close to the magnetic instability and fix $J/U=1/6$ as it has been used earlier for FeSe\cite{Kreisel2017}.
Within this current Ansatz, the orbital susceptibility in the normal state is given by
\begin{align}
	\tilde\chi_{\mu_1 \mu_2 \mu_3 \mu_4}^0 (q) & = - \!\!\!\sum_{k,m,m' }\!\! M_{\mu_1 \mu_2 \mu_3 \mu_4}^{mm'} (\k,\q)  G^{m} (k+q)  G^{m'} (k),   \label{eqn_supersuscept}
\end{align}
where we have adopted the shorthand notation $k\equiv (\k,\omega_n)$ and defined the abbreviation
\begin{eqnarray}
	 M_{\mu_1 \mu_2 \mu_3 \mu_4}^{mn} (\k,\q)& =&\sqrt{Z_{\mu_1}Z_{\mu_2}Z_{\mu_3}Z_{\mu_4}}   \\&&\!\!\times a_n^{\mu_4} (\k) a_n^{\mu_2,*} (\k) a_m^{\mu_1} (\k\!+\!\q) a_m^{\mu_3,*} (\k\!+\!\q).\nonumber
\end{eqnarray}
To evaluate the susceptibility, we perform the frequency summation analytically and numerically sum over the full Brillouin zone using $400\times 400$ $k$ points. Note that the susceptibility is just related by 
\begin{equation}
 \tilde \chi_{\mu_1 \mu_2 \mu_3 \mu_4}^0 (\q)=\sqrt{Z_{\mu_1}Z_{\mu_2}Z_{\mu_3}Z_{\mu_4}}\;\chi_{\mu_1 \mu_2 \mu_3 \mu_4}^0 (\q),
 \label{eq_susc}
\end{equation}
to the one from a fully coherent electronic structure, i.e. the quasiparticle weights enter as prefactors and renormalize each component of the susceptibility tensor.
Finally, we treat the interactions in a random phase approximation (RPA) to calculate the spin (1) and charge (0) susceptibilities
\begin{subequations}
\label{eqn:RPA}
\begin{align}
 \tilde\chi_{1\,\mu_1\mu_2\mu_3\mu_4}^{\rm RPA} (\q) &= \left\{\tilde \chi^0 (q) \left[1 -\bar U^s \tilde\chi^0 (q) \right]^{-1} \right\}_{\mu_1\mu_2\mu_3\mu_4},\\
 \tilde\chi_{0\,\mu_1\mu_2\mu_3\mu_4}^{\rm RPA} (\q) &= \left\{\tilde \chi^0 (q) \left[1 +\bar U^c \tilde\chi^0 (q) \right]^{-1} \right\}_{\mu_1\mu_2\mu_3\mu_4}.
\end{align}
\end{subequations}
The total spin susceptibility as measured experimentally is given by the sum
\begin{equation}
	\label{eqn_chisum}\tilde \chi (\q,\omega) = \frac 12 \sum_{\mu \nu} \tilde\chi_{1\;\mu \mu \nu \nu}^{\rm RPA} (\q,\omega)\,.
 \end{equation}
Note that the interaction matrices $\bar U^s$ and $\bar U^c$ contain linear combinations of the parameters $U,U',J,J'$, for details see for example Ref.~\onlinecite{a_kemper_10}.

To calculate the superconducting instability in the spin-singlet channel (the dominant one for the present models), we use the  vertex for pair scattering between bands $n$ and $m$,
 \begin{align}
	&\tilde{\Gamma}_{n m} (\mathbf{k},\mathbf{k}')  = \mathrm{Re}\sum_{\mu_1 \mu_2 \mu_3 \mu_4} a_{n}^{\mu_1,*}(\mathbf{k}) a_{n}^{\mu_4,*}(-\mathbf{k}) \nonumber\\&\hspace{0.6cm}\times 
	{\tilde\Gamma}_{\mu_1\mu_2\mu_3\mu_4} (\k,\k')
	 a_{m}^{\mu_2}(\mathbf{k}') a_{m}^{\mu_3}(-\mathbf{k}')\, \label{eq_Gam_mu_nu}\,,
\end{align}
where $\k$ and $\k'$ are momenta restricted to the pockets $\k \in C_n$ and $\k' \in C_m$, and is defined in terms of the  
  the orbital space vertex function 
\begin{align}
	&{\tilde\Gamma}_{\mu_1\mu_2\mu_3\mu_4} (\k,\k') = \left[\frac{3}{2} \bar U^s \tilde\chi_1^{\rm RPA} (\k-\k') \bar U^s \right.\,\,\\
	&\,\left. +  \frac{1}{2} \bar U^s - \frac{1}{2}\bar U^c \tilde\chi_0^{\rm RPA} (\k-\k') \bar U^c + \frac{1}{2} \bar U^c \right]_{\mu_1\mu_2\mu_3\mu_4}\nonumber \label{eq:fullGammadress}
\end{align}
Then, the linearized gap equation
 \begin{equation}\label{eqn:gapeqn}
-\frac{1}{V_G}  \sum_m\int_{\text{FS}_m}dS'\; \tilde\Gamma_{nm}(\k,\k') \frac{ g_i(\k')}{|v_{\text{F}m}(\k')|}=\lambda_i g_{i}(\k)
 \end{equation}
describes the superconducting gap $\Delta(\k)\propto g(\k)$ for the largest eigenvalue $\lambda$ at least at $T_c$. 
The integration is over the Fermi surface $\text{FS}_m$, the Fermi velocity $v_{\text{F}m}(\k')$ enters as weights in the denominator and $V_G$ is the volume of the Brillouin zone. For the uncorrelated case (Figs. \ref{fig:chi}, \ref{fig_gap} (a)) we have used $Z_\alpha=1$, while for the weakly correlated case we start with $\sqrt{Z_\alpha}= [0.72,
0.89,
0.77,
0.77,
0.85
]$
% real numbers are those: 0.72111, 0.89443, 0.7746,0.7746, 0.84853
\cite{Bjornson_2020}, subsequently reduce $\sqrt{Z_{xy}}$ by steps of $0.1$, Figs. \ref{fig:chi}, \ref{fig_gap}(b), and finally split $\sqrt{Z_{xz,yz}}=0.77\mp 0.02s$, $s=1,2,3,4$, Figs. \ref{fig:chi}, \ref{fig_gap}(c).

% \bibliography{references_noYpocket}
%apsrev4-2.bst 2019-01-14 (MD) hand-edited version of apsrev4-1.bst
%Control: key (0)
%Control: author (8) initials jnrlst
%Control: editor formatted (1) identically to author
%Control: production of article title (0) allowed
%Control: page (0) single
%Control: year (1) truncated
%Control: production of eprint (0) enabled
%

\end{document}